\documentclass[a4paper,11t]{article}
\usepackage[top=2.54cm,bottom=2.54cm, left=2cm, right=2cm]{geometry}
\usepackage{graphicx} 
\usepackage{booktabs}
\usepackage{tabu}
\usepackage{amsmath,amssymb}
\usepackage{float}
\usepackage{alphabeta}
\usepackage{multirow}
\usepackage{bm}
\usepackage{enumitem}
\usepackage[colorlinks=true, allcolors=blue]{hyperref}
\usepackage[export]{adjustbox}
\usepackage[sort,comma]{natbib}
\usepackage{comment}
\usepackage{rotating} 

\title{Simplicial clustering using the $\alpha$--transformation}
\author{Michail Tsagris, Christos Adam and Nikolaos Kontemeniotis  \\
\\
Department of Economics, University of Crete, Greece  \\
\href{mailto:mtsagris@uoc.gr}{mtsagris@uoc.gr},  
\href{mailto:econp266@econ.soc.uoc.gr}{econp266@econ.soc.uoc.gr} and
\href{mailto:kontemeniotisn@gmail.com}{kontemeniotisn@gmail.com} \\
}

\begin{document}

\maketitle

\begin{center}
\textbf{Abstract}   \\
\end{center}
We introduce two simplicial clustering approaches for compositional data, that are adaptations of the $K$--means and of the Gaussian mixture models algorithms, by employing the $\alpha$--transformation. By utilizing clustering validation indices we can decide on the number of clusters and choose the value of $\alpha$ for the $K$--means, while for the model-based clustering approach information criteria complete this task. Extensive simulation studies compare the performance of these two approaches and a real dataset illustrates their performance in real world settings.  

\section{Introduction}
Compositional data are non-negative multivariate data which sum to the same constant, usually $1$, and in this case, their sample space is the standard simplex
\begin{eqnarray} \label{simplex}
\mathbb{S}^{p-1}=\left\lbrace(x_1,...,x_p)^\top \bigg\vert x_i \geq 0,\sum_{i=1}^px_i=1\right\rbrace, 
\end{eqnarray}
where $D$ denotes the number of variables (better known as components). 

Compositional data are met in many different scientific fields. See \cite{tsagris2020folded} for a collection of cases illustrating the breadth of compositional data analysis applications and consequently the need for simplicial models and algorithms.

The natural question that arises is how to account for the compositional constraint. A simple approach is to ignore the compositional constraint and treat the data as though they were Euclidean, an approach termed "Euclidean data analysis" (EDA) \citep{baxter2001, baxter2005,baxter2006, woronow1997elusive, woronow1997regression}. \cite{ait1980,ait1982,ait2003} contended that data should instead be analysed after applying logarithm based transformations yielding the log-ratio analysis (LRA) approach, arguing that this amounted to working with an implied distance measure on the simplex (discussed further in the next section) that satisfied particular mathematical properties he regarded as essential for compositional data analysis.   

One characteristic of a dataset that immediately rules out using LRA in its standard form is the presence of observations for which one or more components is zero, since for such observations log-ratio transformations are undefined. Data of this type are not uncommon, so this is a notable weakness of LRA. Some attempts have been made to modify LRA to make it appropriate for data containing zeros (particularly when the zeros are assumed to arise
from rounding error), but these involve a somewhat ad hoc imputation approach of replacing zeros with small values. 

\citet{scealy2014colours} have recently presented a historical summary of the debate, and have given a critical appraisal of the properties often invoked by authors to support the use of LRA. We share Scealy and Welsh's opinion that LRA should not be a default choice for compositional data analysis on account of such properties. In this paper, we take the pragmatic view, which seems especially relevant for clustering problems that we should adopt whichever approach performs best in a given setting. A side message of this paper is that for clustering problems, the choice of whether or not one should transform the data, and if so which transformation to use, should depend on the dataset under study.

There is also a plethora of alternative power transformations for compositional data. \cite{ait2003} defined the Box-Cox transformation applied to ratios of components, while \cite{greenacre2008, greenacre2010} applied the Box-Cox transformation to each of the components, and more recently he introduced another Box-Cox power transformation \citep{greenacre2024}. Another power transformation is the folded-power transformation \citep{atkinson1985}, extended to compositional data. \cite{tsagris2011} proposed the $\alpha$--transformation, a generalisation of the isometric log-ratio transformation \cite{ilr2003}, and finally, \cite{clarotto2022} proposed a modification of the $\alpha$--transformation, termed the isometric $\alpha$--transformation. One last transformation, that does not include a power-parameter though, is the square root transformation \citep{stephens1982,scealy2011a,scealy2011b}.

Clustering is an important task in statistics, machine learning and data science in general. Specifically to compositional data, numerous applications can be found. \cite{korhovnova2009} performed hierachical cluster analysis to different types of coffee, including both, Arabica, Robusta and various. \cite{bruno2011} performed functional clustering to trajectories of compositional data to cluster particulate matter vertical profiles in the lower troposphere. \cite{zhou2018} performed cluster analysis using hierarchical clustering, $K$--means and fuzzy c-means to geochemical data that describe stream sediment samples, after transforming the compositional data using various transformations. \cite{lu2018}, \cite{godichon2019}, and \cite{muliawati2023} performed $K$--means to raw meal composition, transcriptomic data \& bicycle sharing data, and to the geochemical compositions of wood fossils, respectively. \cite{cicchella2022} employed model-based clustering and model-free algorithms to establish geochemical backgrounds in stream sediments of an onshore oil deposits area.

Simplicial clustering algorithms have been proposed in the recent years. \cite{cao2013} developed a spatial clustering approach to discover interesting regions in cities in the space of spatial attributes, forming clusters based on different building type signatures. \cite{wang2021a} and \cite{wang2021b} penalized the minimization function of the $K$--means algorithm using the $L_1$ norm and a combination of an $L_2$-norm and an $L_1$-norm regularization, respectively. 

A drawback of the $K$--means is that it assumes spherical clusters of similar size, which may not suit all data types. It is sensitive to initial centroids and may converge to a local minimum. Additionally, the value of $k$ must be chosen in advance, and clustering validity indices (CVIs) are often used to choose the suitable value of $K$. On the other hand, $K$--means is a simple yet powerful clustering technique. 

Simplicial model-based clustering has also been studied, with the first approach relying on the EDA \citep{papageorgiou2001} employing a Gaussian mixture model. Other approaches rely on generalised Dirichlet mixture models \citep{bouguila2009}, structured Dirichlet mixture models \citep{migliorati2017}, Dirichlet mixture models (DMMs), \citep{pal2022}, and Dirichlet-tree multinomial mixtures \citep{mao2022}. A drawback of model-based clustering is that it is parametric and assumes specific, typically ellipsoidal-like, shapes on the clusters. On the other hand, it is more general than the $K$--means that assumes spherical shaped clusters and hence can be seen as a special case of model-based clustering.

In this paper we propose the use of the flexible $\alpha$--transformation prior to the application of $K$--means or model-based clustering, yielding the $\alpha$--$K$--means algorithm. To choose the number of clusters produced by $K$--means, numerous CVIs have been proposed over the years. Among them, we selected 33 indices to investigate. These indices will further allow us to choose the optimal value of $\alpha$. The use of the $\alpha$--transformation will be further coupled with all 14 Gaussian parsimonious clustering models GPCMs \citep{celeux1995,browne2014,mcnicholas2016}, where the choice of the optimal number of clusters and most suitable Gaussian model is performed via the Bayesian information criterion (BIC) \citep{schwarz1978estimating}. This approach will be termed $\alpha$--GPCM.

Extensive experiments using and simulated datasets will examine the performance of either approach. The studies will facilitate an inter-comparison between the two approaches in terms of a) accuracy of selecting the appropriate number of clusters, and b) computational cost. The studies will further allow for an intra-comparison of the 33 CVIs when used with compositional data. A real data analysis will illustrate the performance of the $\alpha$--$K$--means and of the $\alpha$--GPCM.  

The paper is structured as follows. In Section \ref{sec:alpha} we describe the $\alpha$--transformation, in Sections \ref{sec:kmeans} and \ref{sec:gpcm} we describe the $K$--means and model-based clustering, respectively, and combine the two approaches with the $\alpha$--transformation to develop simplicial clustering algorithms for compositional data. Section \ref{sec:alt} delineates alternative approaches for simplicial clustering. Section \ref{sec:sim} contains extensive simulation studies comparing the two proposed simplicial clustering methods and Section \ref{sec:real} illustrates their performance using real data. Section \ref{Sec:conc} concludes the paper. 

\section{The $\alpha$--transformation}  \label{sec:alpha}
For a composition $\bm{x} \in \mathbb{S}^{p-1}$, the centered log-ratio transformation is defined in Aitchison (1983) as 
\begin{eqnarray} \label{clr}
\bm{w}_0(\bm{x})=\left ( \log\left ({\frac{x_1}{\prod_{j=1}^px_j^{1/p}}} \right ),\ldots, \log\left ({\frac{x_p}{\prod_{j=1}^Dx_j^{1/p}}} \right ) \right ).
\end{eqnarray}
The sample space of (\ref{clr}) is the set
\begin{eqnarray} \label{Qp}
\mathbb{Q}_0^{p-1}=\left\lbrace \left(w_{1,0}, \ldots, w_{p,0} \right)^\top: \sum_{i=1}^pw_{i,0}=0 \right\rbrace,
\end{eqnarray}
which is a subset of $\mathbb{R}^{p-1}$. Note that the zero sum constraint in Equation (\ref{Qp}) is an obvious drawback of this transformation as it can lead to singularity issues. In order to remove the redundant dimension imposed by this constraint, one can apply the isometric log-ratio transformation 
\begin{eqnarray} \label{ilr}
\bm{y}_0(\bm{x})=\bm{H}\bm{w}_0(\bm{x}),
\end{eqnarray}
where $\bm{y}_0(\bm{x})$ is a $p-1$ dimensional vector and $\bm{H}$ is the $(p-1) \times p$ Helmert \citep{helm1965} sub-matrix\footnote{This is the Helmert matrix after deletion of the first row. This sub-matrix is a standard orthogonal matrix in shape analysis used to overcome singularity problems. Its structure and components can be found in \citep{tsagris2020folded} and for further information, see \cite{dryden1998,le1999}.}. The sample space of (\ref{ilr}) is $\mathbb{R}^{p-1}$ because left multiplication by the Helmert sub-matrix maps the centered log-ratio transformed data  onto $\mathbb{R}^{p-1}$, thus, in effect, removing the zero sum constraint. 

\cite{tsagris2011} developed the $\alpha$--transformation as a more general transformation than that in Equation (\ref{ilr}). Let 
\begin{eqnarray} 
\label{stayalpha}
\bm{u}_{\alpha}(\bm{x})=\left( \frac{x_1^{\alpha}}{\sum_{j=1}^px_j^{\alpha}}, \ldots, \frac{x_p^{\alpha}}{\sum_{j=1}^px_j^{\alpha}} \right)^\top
\end{eqnarray}
denote the power transformation for compositional data as defined by \cite{ait2003}, where $\alpha$ can take any value\footnote{When zero values exist in the data , $\alpha$ can take only strictly positive values.}. In a manner analogous to Equations (\ref{clr}-\ref{ilr}), first define
\begin{eqnarray} \label{alef}
\bm{w}_{\alpha}(\textbf{x})=\frac{p\bm{u}_{\alpha}(\textbf{x})-1}{\alpha}.
\end{eqnarray}
The sample space of Equation (\ref{alef}) is 
\begin{eqnarray}  \label{Qad}
\mathbb{Q}_{\alpha}^{p-1}=\left\lbrace \left(w_{1, \alpha}, \ldots, w_{p, \alpha} \right)^\top:\frac{-1}{\alpha} \leq w_{i, \alpha} \leq \frac{p-1}{\alpha},\sum_{i=1}^pw_{i,\alpha}=0 \right\rbrace.
\end{eqnarray}
Note that the inverse of Equation (\ref{alef}) is   
\begin{equation}
\label{winv}
\bm{x} =\bm{w}^{-1}_{\alpha}(\bm{w}) = \left (\frac{(1+\alpha w_1)^{1/\alpha}}{\sum_{j=1}^p (1+\alpha w_j)^{1/\alpha}},\ldots,\frac{(1+\alpha w_p)^{1/\alpha}}{\sum_{j=1}^p (1+\alpha w_j)^{1/\alpha}} \right ),
\end{equation}
for $\bm{w} \in \mathbb{Q}_{\alpha}^{p-1}$. As $\alpha \rightarrow 0$ Equation (\ref{alef}) converges to Equation (\ref{clr}) and Equation (\ref{winv}) becomes
\begin{equation}
\label{winv0}
\bm{x} =\bm{w}^{-1}_0(\bm{w}) = \left (\frac{e^ {w_1}}{\sum_{j=1}^p e^{w_j}},\ldots,\frac{e^ {w_p}}{\sum_{j=1}^p e^{w_j}} \right ).
\end{equation}
Finally, the $\alpha$--transformation is defined as
\begin{eqnarray} \label{alpha}
\bm{y}_{\alpha}(\bm{x})=\bm{Hw}_{\alpha}(\bm{x}).
\end{eqnarray}
The transformation in Equation (\ref{alpha}) is a one-to-one transformation which maps data inside the simplex onto a subset of $\mathbb{R}^{p-1}$ and vice versa for 
$\alpha \neq 0$. The corresponding sample space of Equation (\ref{alpha}) is 
\begin{eqnarray} \label{Ap}
\mathbb{A}_{\alpha}^{p-1}=\left\lbrace\bm{H}\bm{w}_{\alpha}(\bm{x}) \bigg | -\frac{1}{\alpha} \leq w_{i,\alpha} \leq \frac{p-1}{\alpha},\sum_{i=1}^pw_{i, \alpha}=0 \right\rbrace.
\end{eqnarray}
For $\bm{y} = \bm{z}_\alpha(\bm{x})$, the inverse transformation from $\mathbb{A}_{\alpha}^{p-1}$ to $\mathbb{S}^{p-1}$ is $\bm{z}_\alpha^{-1}(\bm{y}) = \bm{w}^{-1}_\alpha(\bm{H}^\top\bm{y})$ where $\bm{w}^{-1}(\cdot)$ is given in Equation (\ref{winv}). Note that vectors in $\mathbb{A}_{\alpha}^{p-1}$ are not subject to the sum to zero constraint and that $\lim_{\alpha \rightarrow 0}\mathbb{A}_{\alpha}^{p-1} \rightarrow \mathbb{R}^{p-1}$.

For convenience purposes we allow $\alpha$ to lie within $\left[-1,1\right]$. From Equations (\ref{stayalpha}) and (\ref{alef}), when $\alpha=1$, the simplex is linearly expanded as the values of the components are simply multiplied by a scalar and then centered. When $\alpha=-1$, the inverse of the values of the components are multiplied by a scalar and then centered.

\cite{tsagris2011} argued that while the $\alpha$--transformation did not satisfy some of the properties that \cite{ait2003} deemed important, this was not a downside of this transformation as those properties were suggested mainly to fortify the concept of log-ratio methods. In compositional data the relevant information is contained in the (log-)ratios between the components, thus the desirable properties are perfectly coherent with the nature on these data. All of this makes the logratio methodology a consistent approach. \cite{scealy2014} also questioned the importance of these properties and, in fact, showed that some of them are not actually satisfied by the log-ratio methods that they were intended to justify.

\section{$K$--means}  \label{sec:kmeans}
$K$--means is a popular unsupervised learning algorithm used in clustering analysis. It partitions a given dataset into $K$ distinct, non-overlapping clusters. Each data point belongs to the cluster with the nearest mean, serving as a prototype of the cluster. $K$--Means is widely used in market segmentation \citep{tsiptsis2011}, image compression \citep{zhang1996}, document classification \citep{steinbach2000} and tumor classification \citep{dudoit2002}.

Given a dataset ${\bm X} \in \mathbb{R}^p$ consisting of $n$ observations, and a predefined number of clusters $K$, the $K$--means algorithm proceeds as follows:

\begin{enumerate}
    \item \textbf{Initialization Step:} Choose $K$, the number of initial cluster centroids (or barycenters) $\bm{\mu}_1, \bm{\mu}_2, \dots, \bm{\mu}_K$ randomly.
    \item \textbf{Assignment Step:} Assign each data point $\bm{x}_i$ to the cluster $C_j$ with the nearest centroid
    $$
    C_k = \{\bm{x}_i : \|\bm{x}_i - \bm{\mu}_k\|^2 \leq \|\bm{x}_i - \bm{\mu}_k\|^2, \forall k \in \{1, \dots, K\}\},
    $$
    where $\| . \|$ denotes the Euclidean distance.
    \item \textbf{Update Step:} Recalculate the centroid of each cluster
    $$
    \bm{\mu}_k = \frac{1}{|C_k|} \sum_{\bm{x}_i \in C_k} \bm{x}_i,
    $$
    where $|C_k|$ denotes the cardinality of the set $C_k$.
    \item \textbf{Repeat:} Iterate steps 2 and 3 until convergence (i.e., the assignments no longer change or the centroids stabilize).
\end{enumerate}

The $K$--means algorithm minimizes the within-cluster sum of squares (WCSS), the sum of the distances of all points from the centroids of the clusters to whom they belong $WCSS=\sum_{k=1}^{K} \sum_{\bm{x}_i \in C_k} \|\bm{x}_i - \bm{\mu}_k\|^2$. Effectively, the algorithm aims to find cluster assignments and centroids that minimize the $WCSS$.

\subsection{Cluster validity indices}
Prior to delineating the cluster validity indices (CVIs), let us mention some preliminary quantities. 

The total dispersion is simply the covariance matrix multiplied by $n-1$, 
$$T = \sum_{i=1}^n\left(\bm{x}_i-\bm{\mu}\right)\left(\bm{x}_i-\bm{\mu}\right)^\top.$$ 
The total sums of squares $TSS$ is the trace of this matrix, $TSS=\text{Tr}(T)$. The within-cluster dispersion of cluster $k$ is given by 
$$WC_k=\sum_{k=1}^{K}\left(\bm{x}_i - \bm{\mu}_k\right)\left(\bm{x}_i - \bm{\mu}_k\right)^\top,$$ and subsequently, $WC=\sum_{k=1}^KWC_k$, and evidently, $WCSS = \text{Tr}(WC)$. Finally, the between-cluster (BC) dispersion of cluster $k$ is given by 
$$BC_k=\sum_{k=1}^{K}n_k\left(\bm{\mu}_k - \bm{\mu}\right)\left(\bm{\mu}_k - \bm{\mu}\right)^\top,$$ the between-cluster dispersion is $BC=\sum_{k=1}^{K}BC_k$, and the between-cluster sum of squares is $BCSS=\text{Tr}(BC).$

Let us now define the following distance based quantities,
$$\delta_k=\frac{1}{n_k}\sum_{i\in C_k} \|\bm{x}_i-\bm{\mu}_k\| \ \ \text{and} \ \ \Delta_{kk'}=\|\bm{\mu}_k-\bm{\mu}_{k'}\|.$$

In the cluster $C_k$, there are $n_k$ observations and
$$n_W = \sum_{k=1}^K \frac{n_k(n_k-1)}{2} = \frac{1}{2} \left( \sum_{k=1}^K n_k^2 - n \right)$$
pairs of observations. The total number of pairs of observations is $n_T=\frac{n(n-1)}{2}$. Since $n=\sum_{k=1}^K n_k$, 
$$n_T\ =\ \frac{n(n-1)}{2}  =  \frac{1}{2} \left( \sum_{k=1}^K n_k\right)^2 - \frac{1}{2}\sum_{k=1}^K n_k  =  n_W + \sum_{k<k'}n_kn_{k'} = n_W + n_B,$$
where $n_B = \sum_{k<k'}n_kn_{k'}$ denotes the number of pairs of observations which do not belong to the same cluster.

The subsequent sections contain various CVIs used to determine the optimal number of clusters. We begin with the indices whose minimum value is desirable and continue with the indices whose maximum values are desirable. 

\subsection{CVIs with minimum optimal values}
The following CVIs choose the number of clusters that minimize their value. 

\begin{itemize}
\item The Banfeld-Raftery index \citep{banfield-raftery-93} is 
\begin{equation}
BRI=\sum_{k=1}^K n_k\log\left(\frac{Tr(WC_k)}{n_k}\right).  \label{eq-crit-banfeld-raftery}
\end{equation}

\item The Davies-Bouldin index \citep{davies-bouldin-79} is the mean value, among all clusters, of the maximum of the following quantities:
\begin{equation}
DBI=\frac{1}{K}\sum_{k=1}^K \max_{k'\ne k}\left(\frac{\delta_k+\delta_{k'}}{\Delta_{kk'}}\right).
\label{eq-crit-davies-bouldin}
\end{equation}

\item The Det\_Ratio index \citep{scott-symons-71} is
\begin{equation}
DRI=\frac{\det(T)}{\det(WC)}.   \label{eq-crit-Det-Ratio}
\end{equation}

\item The Log\_Det\_Ratio index \citep{scott-symons-71} is a function of the logarithm of the previous index
\begin{equation}
LDRI=N\,\log\left(\frac{\det(T)}{\det(WC)}\right).    \label{eq-crit-Log-Det-Ratio}
\end{equation}

\item The Log\_SS\_Ratio index \citep{hartigan-75} is 
\begin{equation}
LSSI =\log\left(\frac{BCSS}{WCSS}\right).   \label{eq-crit-Log-SS-Ratio}
\end{equation}

\item The McClain-Rao index \citep{mcclain-rao-01} is
\begin{equation}
MRI=\frac{n_B}{n_W} \frac{\sum_{k=1}^K \sum_{ {i,j\in C_k}\atop {i<j} } \|\bm{x}_i -\bm{x}_j\|}{\sum_{k<k'} \sum_{ {i\in C_k,\,j\in C_{k'}} \atop {i<j}}\|\bm{x}_i -\bm{x}_j\|}.
\label{eq-crit-McClain-Rao}
\end{equation}

\item The Ray-Turi index \citep{ray-turi-99} is 
\begin{equation}
RTI= \frac{1}{n}\frac{WCSS}{\displaystyle\min_{k<k'} \Delta_{kk'}^2}.  \label{eq-crit-Ray-Turi}
\end{equation}

\item The Scott-Symons index \citep{scott-symons-71} is
\begin{equation}
SSI=\sum_{k=1}^K n_k\log\left[\det\left(\frac{WC_k}{n_k}\right)\right].    \label{eq-crit-scott-symons}
\end{equation}

\item The Xie-Beni \citep{xie-beni-91} is 
\begin{equation}
XBI= \frac{1}{n}\frac{WCSS}{\displaystyle\min_{k<k'} \delta_1(C_k,C_{k'})^2},
\label{eq-crit-Xie_Beni}
\end{equation}
where $$ \delta_1(C_k,C_{k'}) = \min_{ {i\in C_k} \atop {j\in C_{k'}}}\|\bm{x}_i-\bm{x}_j\|. $$

\end{itemize}

\subsection{CVIs with maximum optimal values}
The following CVIs choose the number of clusters that maximize their value.

\begin{itemize}

\item The Ball-Hall index \citep{ball-hall-65} is 
\begin{equation}
BHI=\frac{1}{K}\sum_{k=1}^K \frac{1}{n_k}\sum_{i\in I_k} \|\bm{x}_i-\bm{\mu}_k\|^2.
\label{eq-crit-ball-hall}
\end{equation}

\item The Calinski-Harabasz index \citep{calinski-harabasz-74} is 
\begin{equation}
CHI=\frac{BCSS/(K-1)}{WCSS/(n-K)}=\frac{n-K}{K-1}\frac{BCSS}{WCSS}.  \label{eq-crit-calinski}
\end{equation}

\item The distance between clusters $C_k$ and $C_{k'}$ is measured by the distance between their closest points
$$ d_{kk'}=\min_{{i\in C_k}\atop{j\in I_{k'}}} \|\bm{x}_i^k-\bm{x}_i^k\| $$
and $d_{min}$ is the smallest of these distances $d_{kk'}$, $d_{min}=\min_{k\ne k'}d_{kk'}$. Similarly, for each cluster $C_k$, let the $D_k$ denotes the largest distance separating two distinct points in the cluster $$ D_{kk'}=\max_{{i,j\in I_k}\atop{i\ne j}} |\|\bm{x}_i^k-\bm{x}_i^k\|. $$ The $d_{max}$ is the largest of these distances $D_k$, $d_{max}=\max_{1\leq k\leq K}D_k$. 

The Dunn index \citep{dunn-74} is then defined as
\begin{equation}
DI=\frac{d_{min}}{d_{max}}.   \label{eq-crit-dunn}
\end{equation}

\item To define the generalised Dunn (GD) indices, let us denote by $\delta$ a measure of the between-cluster distance and by $\Delta$ a measure of the within-cluster distance. Five\footnote{In fact there are six, but we have skipped the last one.} different definitions of $\delta$ (denoted as $\delta_1$ through $\delta_{6}$) and three definitions of $\Delta$ (denoted as $\Delta_{1}$ through $\Delta_{3}$) have been suggested. This leads to 15 different indices. 

The definitions of the three within-cluster distances $\Delta$ are:
\begin{eqnarray*}
    \Delta_{1}(C_k) & = & \max_{{i,j\in C_k}\atop {i\ne j}}\|\bm{x}_i-\bm{x}_j\| \\
    \Delta_{2}(C_k) & = & \frac{1}{n_k(n_k-1)}\sum_{ {i,j\in C_k}\atop {i\ne j} }\|\bm{x}_i-\bm{x}_j\| \\
    \Delta_{3}(C_k) & = & \frac{2}{n_k}\sum_{i\in C_k}\|\bm{x}_i-\bm{\mu}_k\|.
\end{eqnarray*}

The definitions of the five between-cluster distances $\delta$ are:
\begin{eqnarray*}
    \delta_{1}(C_k,C_{k'}) & = & \min_{ {i\in I_k} \atop {j\in I_{k'}}}\|\bm{x}_i-\bm{x}_j\| \\
    \delta_{2}(C_k,C_{k'}) & = & \max_{ {i\in I_k} \atop {j\in I_{k'}}}\|\bm{x}_i-\bm{x}_j\| \\
    \delta_{3}(C_k,C_{k'}) & = & \frac{1}{n_kn_{k'}}\sum_{ {i\in I_k} \atop {j\in I_{k'}}}\|\bm{x}_i-\bm{x}_j\| \\
    \delta_{4}(C_k,C_{k'}) & = &  \|\bm{\mu}_i - \bm{\mu}_{k'}\| \\
    \delta_{5}(C_k,C_{k'}) & = & \frac{1}{n_k+n_{k'}} 
    \left(\sum_{i\in C_k}\|\bm{x}_i - \bm{\mu}_k\| + \sum_{j\in 
    C_{k'}}\|\bm{x}_i - \bm{\mu}_{k'}\| \right).
\end{eqnarray*}

The GD indices \citep{bezdek-pal-98} are generalisations of the Dunn index \citep{dunn-74} defined as
\begin{equation}
GDI=\frac{\min_{k\ne k'}\delta(C_k,C_{k'})}{\max_k\Delta(C_k)}.
\label{eq-crit-GDI}
\end{equation}

\item The Ksq\_DetW index \citep{marriot-75}, also denoted as $k^2\,|W|$, is
\begin{equation}
KWI=K^2\det(WC)   \label{eq-crit-ksq-detW}.
\end{equation}

\item The PBM index\footnote{The acronym is formed by the initials of the surnames of the authors, Pakhira, Bandyopadhyay and Maulik.} \citep{pakhira-04} is 
\begin{equation}
PBMI= \left( \frac{1}{K}\times\frac{\sum_{i=1}^n\|\bm{x}_i - \bm{\mu}\|}{\sum_{k=1}^K\sum_{i\in I_k}\|\bm{x}_i - \bm{\mu}_k\|}\times \max_{k<k'} \|\bm{\mu}_i - \bm{\mu}_{k'}\| \right)^2.  \label{eq-crit-PBM}
\end{equation}

\item The point-biserial index \citep{milligan-81} is 
\begin{equation}
PBI=\left(\frac{\sum_{k=1}^K \sum_{ {i,j\in C_k}\atop {i<j} } \|\bm{x}_i -\bm{x}_j\|}{n_W}-\frac{\sum_{k<k'} \sum_{ {i\in C_k,\,j\in C_{k'}} \atop {i<j}}\|\bm{x}_i -\bm{x}_j\|}{n_B}\right)\frac{\sqrt{n_Wn_B}}{n_T}.
\label{eq-crit-point-biserial}
\end{equation}

\item The Ratkowsky-Lance index \citep{ratkowsky-lance-78} is 
\begin{equation}
RLI=\sqrt{\frac{\frac{1}{p}\sum_{j=1}^p \frac{BCSS_j}{TSS_j}}{K}},   \label{eq-crit-cbar-k}
\end{equation}
where 
\begin{eqnarray*}
BCSS_j = \sum_{k=1}^K n_k(\bm{\mu}_k-\bm{\mu}_j)^2  \ \ \text{and} \ \ 
TSS_j = \sum_{i=1}^n(x_{ij}-\bm{\mu}_j)^2.
\end{eqnarray*}

\item For the silhouette index let us consider, for each point $\bm{x}_i$, its mean distance to each cluster. One defines the within-cluster mean distance $a(i)$ as the mean distance of point $\bm{x}_i$ to the other points of the cluster it belongs to 
\begin{equation}
a(i) = \frac{1}{n_k-1}\sum_{ {i, j\in C_k} \atop {i\ne j}} \|\bm{x}_i - \bm{x}_j\|.
\end{equation}

Compute the mean distance of $x_i$ to the points of each of the other clusters $C_{k'}$ and denote by $b(i)$ the smallest of these mean distances
\begin{equation}
b(i) = \min_{k'\ne k} \left\lbrace \frac{1}{n_{k'}}\sum_{j\in C_{k'}} \|\bm{x}_i - \bm{x}_j\| \right\rbrace.
\end{equation}

The value of $k'$ which minimizes $b(i)$ denotes the cluster to which the observation $\bm{x}_i$ could probably fit better than the cluster it was found to belong to by the clustering algorithm.

The silhouette \citep{rousseeuw-87} of the observation $\bm{x}_i$ is then computed as 
\begin{equation}
s(i) = \frac{b(i)-a(i)}{\max\left\lbrace a(i),b(i)\right\rbrace}.
\end{equation}

Finally, the silhouette index is the mean of the mean silhouettes through all the clusters:
\begin{equation}
SI=\frac{1}{K}\sum_{k=1}^K \frac{1}{n_k}\sum_{i\in C_k} s(i).   \label{eq-crit-Silhouette}
\end{equation}
 
\item The Trace\_WiB (or Trace\_$W^{-1}B$) index \citep{friedman-rubin-67} is 
\begin{equation}
TWBI=\text{Tr}(WC^{-1}BC).    \label{eq-crit-Trace-WiB}
\end{equation}

\end{itemize}

\subsection{Comments on the CVIs}
\cite{chouikhi2015comparison} showed that the LSSI and the SI indices are the most effective, correctly identifying the number of clusters in over 87\% of runs. TWBI and CHI indices also show strong performance at 86\%. Overall, CHI, and DBI, among others, emerge as the most reliable indices for estimating the correct number of clusters in artificial datasets, highlighting their robustness across different clustering scenarios artificial datasets.

\cite{olatz2013} compared a wide range of CVIs using a corrected evaluation methodology. While no index consistently outperforms all others, the Silhouette index frequently achieves strong results. Noise and cluster overlap have the most significant negative impact, greatly reducing index accuracy. Interestingly, some indices perform better in complex scenarios, such as higher dimensionality or non-uniform cluster densities. Statistical analysis reveals three main performance groups, with DBI, CHI, GDI and SI among the top performers. 
 
\cite{todeschini2024} evaluated 68 CVIs across 21 benchmark datasets with known true cluster numbers. Several indices consistently performed well, including DBI, LSSI, two GDIs, and TWBI. Traditional indices like XBI and SI also performed reliably, with SI showing stable trends across datasets. Indices based on absolute or first min/max rules were found to be more reliable for identifying the natural number of clusters compared to those using max ratio rules, which are more sensitive to small changes in partition values. Only a minority of indices demonstrated full invariance to changes in the maximum number of clusters, with better stability observed in min/max rule-based indices.

\cite{vendramin2010} surveyed and evaluated 40 CVIs, analyzing their computational complexity and performance across nearly one million partitions generated from five clustering algorithms on over a thousand datasets. The indices are categorized into optimization-like and difference-like criteria, with a proposed method to convert between the two. An alternative evaluation methodology is introduced, aiming to address limitations in traditional comparison methods by moving beyond simple accuracy assumptions. Experimental results show that PBMI, PBI, and SI generally offer the most robust performance, with Silhouette standing out for its consistency across different scenarios. Simplified versions of SI perform comparably and may be preferred for large datasets due to lower computational cost. However, some indices show sensitivity to factors like the number of attributes or clusters. The findings are based on datasets with well-formed, normally distributed clusters, so conclusions may not generalize to more complex or irregular data types.

\cite{ikotun2025} focused on internal CVIs used as fitness functions in meta-heuristic-based clustering algorithms. These indices assess clustering quality based on properties like cohesion, separation, symmetry, and connectedness. Among them, the DBI index is the most commonly used However, the DBI performs poorly on datasets with irregular shapes or varying densities. The choice of distance metric affects the shape of clusters that can be detected—Euclidean favors spherical shapes, while other measures like maximum edge or cosine distance capture more complex structures. CVI performance also depends on the clustering method, data structure, and noise level. 

\subsection{The $\alpha$--$K$--means algorithm}
The approach we propose is straightforward, we $\alpha$--transform the data, scale the data, so that all variables have a zero mean and unit variance and then apply the $K$--means algorithm. We shall term this approach $\alpha$--$K$--means. The scaling step is important for two reasons, at first to ensure the robustness of the $K$--means algorithm, since the Euclidean distance utilized by the algorithm is affected by different units of measurements in the data and secondly because different values of $\alpha$ could lead to different ranges in the transformed data. Thus, in this way, the results produced by different values of the $\alpha$ parameter will be comparable. The minimized/maximised values of the CVIs will be a key driver to select both the optimal number of clusters, and to select the optimal value of $\alpha$. 

The significant advantage of this approach is that, apart from the flexibility inherit in the $\alpha$--transformation (\ref{alpha}), we escape the limitation of the isometric log-transformation (\ref{ilr}) which is that the $\alpha$--transformation is applicable even in the presence of zero values. Zero imputation strategies \citep{ait2003,palarea2007} must be applied prior to the isometric (or any other log-ratio related transformation). This approach however induces bias in the data \citep{tsagris2015a}. On the contrary, the $\alpha$--transformation is well defined, but for positive values of $\alpha$ only.

\section{Model-based clustering}  \label{sec:gpcm}
Under a model-based clustering point of view, we model the observed data using GPCMs. The EM algorithm is a standard approach in estimating finite mixture models \citep{mclachlan2000, mcnicholas2016, fruhwirth2019, yao2024}. 

Assume for instance that there are $K\geqslant 1$ underlying clusters in the population with (unknown) weights equal to $q_1,\ldots,q_K$, where $q_j>0$ and $\sum_{j=1}^{K}q_j=1$. Define the (latent) allocation variables $\bm{Z}_i = (Z_{i1},\ldots,Z_{in})^\top$, $i=1,\ldots,n$, with $\bm{Z}_i \sim\mathcal M(1;p_1,\ldots,q_K)$, where the latter represents the multinomial distribution with $K$ categories and success probabilities equal to $q_1,\ldots,q_K$. The prior probability of selecting an observation from cluster $j$ is equal to $\mathrm{P}(Z_{ij} = 1) = q_j$ independent for  $i = 1,\ldots,n$. Conditional on $\bm{Z}_i = \bm{z}_i$, $\bm{Y}_i$ is distributed as
\begin{equation}
\bm{Y}_i| (\bm{Z}_i = \bm{z}_i) \sim f_{\bm{Y}_i| \bm{Z}_i}(\bm{y}_i|\bm{z}_i) =\prod_{k=1}^{K} f^{z_{ik}}(\bm{y}_i;\bm{\mu}_k, \bm{\Sigma}_k),\quad\mbox{independent for}\quad i = 1,\ldots,n
\end{equation}
where $f(\cdot;\bm{\mu}_k, \bm{\Sigma}_k)$ denotes the probability density function of the Gaussian distribution with parameters $\bm{\mu}_k$ and $\bm{\Sigma}_k$ for $k=1,\ldots,K$. Notice that the previous equation in case where the $k$-th element of $\bm{Z}_i$ is equal to 1, collapses to $\bm{Y}_i| (Z_{ik} = 1) \sim f(\cdot;\bm{\mu}_k, \bm{\Sigma}_k)$.
It follows that the joint distribution of $\bm{Y}_i, \bm{Z}_i$ is 
\begin{equation}
\label{eq:joint}
    (\bm{Y}_i, \bm{Z}_i) \sim f_{\bm{Y}_i, \bm{Z}_i}(\bm{y}_i, \bm{z}_i) = \prod_{k=1}^K\left\{q_kf(\bm{y}_i;\bm{\mu}_k, \bm{\Sigma}_k)\right\}^{z_{ik}}.
\end{equation}
Consequently, the marginal distribution of $\bm{Y}_i$ is a mixture of $K$ Gaussian distributions
\begin{equation}
\label{eq:fmm}
    \bm{Y}_i \sim f_{\bm{Y}_i}(\bm{y}_i) =\sum_{k=1}^Kq_k f(\bm{y}_i;\bm{\mu}_k, \bm{\Sigma}_k),\quad\mbox{independent for}\quad i = 1,\ldots,n.
\end{equation}
The probability that observation $i$ is assigned to group $j$, conditional on $\bm{Y}_i = \bm{y}_i$, is equal to 
\begin{equation}
\label{eq:post}
\mathrm{P}(Z_{ik} = 1|\bm{Y}_i = \bm{y}_i) =  w_{ij} = \frac{q_k f\left(\bm{y}_i;\bm{\mu}_k, \bm{\Sigma}_k\right)}{\sum_{\ell=1}^{K}q_j f\left(\bm{y}_i;\bm{\mu}_j, \bm{\Sigma}_j\right)},\quad k=1,\ldots,K
\end{equation}
or equivalently, $\bm{Z}_i|(\bm{Y}_i = \bm{y}_i)\sim\mathcal M(1,w_{i1},\ldots,w_{iK})$, independent for $i=1,\ldots,n$. The observed log-likelihood is defined as 
\begin{equation}
\label{eq:mixlik}
\ell(\bm{\mu}, \bm{\Sigma}, \bm{q}|\bm{y}) =  \sum_{i=1}^{n}\log\left\{\sum_{k=1}^{K}q_k f(\bm{y}_i;\bm{\mu}_k, \bm{\Sigma}_k)\right\},
\end{equation}
where $\bm{\mu} = (\bm{\mu}_{1},\ldots,\bm{\mu}_K)\in\mathbb{R}^{Kp}$ and $\bm{q} = (q_1,\ldots,q_K)\in\mathcal Q_{K-1}$, where $\mathcal Q_{K-1} = \{q_1,\ldots,q_K: q_k > 0, k=1,\ldots,K, \sum_{k=1}^{K}q_k = 1\}$.
In order to maximize \eqref{eq:mixlik}, the EM algorithm is implemented. In brief, given a set of starting values, the algorithm proceeds by computing the expectation of the complete log-likelihood (E-step) and then maximizing with respect to $\bm{\mu}$, $\bm{q}$ (M-step). Consider that we have at hand the \textit{complete} data $(\bm{Y}_i,\bm{Z}_i)$, independent for $i=1,\ldots,n$. From Equation \eqref{eq:joint},  the logarithm of the \textit{complete likelihood} is defined as
\begin{equation}
\label{eq:complete_lik}
\ell^{c}(\bm{\mu}, \bm{\Sigma}_k, \bm{q}|\bm{y},\bm{z}) = \sum_{i=1}^{n}\sum_{k=1}^Kz_{ik}\left\lbrace\log q_k+\log f(\bm{y}_i;\bm{\mu}_k, \bm{\Sigma}_k)\right\rbrace.
\end{equation}
Of course, we do not directly observe $\bm{Z}_i$, $i = 1,\ldots,n$, thus, we cannot use \eqref{eq:complete_lik}. Nevertheless, we can consider the expectation of the logarithm of the complete likelihood with respect to the conditional distribution of $\bm{Z}|\bm{y}$ and a current estimate of $\bm{\mu}$, $\bm{\Sigma}_k$, and $\bm{q}$
\begin{equation}
\label{eq:expected_complete}
\mathrm{E}_{\bm{Z}|\bm{y}} \ell^{c}(\bm{\mu}, \bm{\Sigma}_k, \bm{q}|\bm{y},\bm{Z}) = \sum_{k=1}^Kw_{\cdot k}\log q_k + \sum_{k=1}^K\sum_{i=1}^{n}w_{ik}\log f(\bm{y}_i;\bm{\mu}_k, \bm{\Sigma}_k),
\end{equation}
where $w_{\cdot j} := \sum_{i=1}^{n}w_{ij}$ as defined in \eqref{eq:post}.

In the M-step, we maximize \eqref{eq:expected_complete} with respect to the parameters $\bm{\mu},  \bm{\Sigma}, \bm{p}$. It is easy to see that the mixing proportions are set equal to 
$$ q_k = \frac{w_{\cdot k}}{n},\quad k= 1,\ldots,K.$$ 
Maximization of \eqref{eq:expected_complete} is available in closed form for $\bm{\mu}$.

Let now $\hat{\bm{q}}$, $\hat{\bm{\mu}}$, and $\hat{\bm{\Sigma}}$ denote the estimates of mixing proportions and group-specific parameters, obtained at the last iteration of the EM algorithm. Define the corresponding estimates of the posterior membership probabilities as 
$$\hat w_{ik} = \frac{\hat q_j f(\bm{y}_i;\hat{\bm{\mu}}_k, \hat{\bm{\Sigma}}_k)}{\sum_{j=1}^{K}\hat{q}_j f(\bm{y}_i;\hat{\bm{\mu}}_j, \hat{\bm{\Sigma}}_j)}, \quad k=1,\ldots,K. $$
Then, the resulting single best clustering $\{c_1,\ldots,c_n\}$ arises after applying the Maximum A Posteriori (MAP) rule in the estimated posterior membership probabilities, that is,
\begin{equation*}
c_i = \mathrm{argmax}_k\{\hat w_{ik}, k=1\ldots,K\},\quad i = 1,\ldots,n.
\end{equation*}

Initialization of the EM algorithm demands extra care to avoid convergence to local maxima. Some indicative works on this topic include \cite{biernacki2003choosing, karlis2003choosing, fraley2005incremental, baudry2015mixtures, papastamoulis2016estimation, michael2016effective}. Finally, in order to select the optimal number of clusters and most appropriate model we used the BIC.

\subsection{Fourteen GPCMs}
All the 14 GPCMs \citep{celeux1995} relate to the covariance matrix of each group, which can be writen as
\begin{eqnarray} \label{sigmak}
\bm{\Sigma}_k=\lambda_kD_kA_kD_k^\top, \ \ k=1,\ldots,K,    
\end{eqnarray}
where $\lambda_k=\left|\bm{\Sigma}_k\right|^{1/p}$, $D_k$ is the matrix of eigenvectors of $\bm{\Sigma}_k$, and $A_k$ is a diagonal matrix with the, corresponding, normalized eigenvalues of $\bm{\Sigma}_k$, such that $\left|A_k\right|=1$. Table \ref{gpcms} contains the covariance parameterizations that yield these models and $\bm{I}_p$ denotes the $p$-dimensional identity matrix.
 
\begin{table}[htbp]
\centering
\renewcommand{\arraystretch}{1.3}
\caption{The 14 GPCMs with their relevant covariance structure and a short interpretation.}
\label{gpcms}
\begin{tabular}{l|l|l}
\toprule
\textbf{Model} & \textbf{Covariance Structure} & \textbf{Interpretation} \\
\midrule
EII & $\Sigma_k = \lambda I_p$ & Spherical, equal volume \\
VII & $\Sigma_k = \lambda_k I_p$ & Spherical, variable volume \\ \hline
EEI & $\Sigma_k = \lambda A$ & Diagonal, equal volume \& shape \\
VEI & $\Sigma_k = \lambda_k A$ & Diagonal, variable volume; equal shape \\
EVI & $\Sigma_k = \lambda A_k$ & Diagonal, equal volume; variable shape \\
VVI & $\Sigma_k = \lambda_k A_k$ & Diagonal, variable volume \& shape \\ \hline
EEE & $\Sigma_k = \lambda D A D^T$ & Ellipsoidal, equal volume, shape \& orientation \\
VEE & $\Sigma_k = \lambda_k D A D^T$ & Ellipsoidal, variable volume; equal shape \& orientation \\
EVE & $\Sigma_k = \lambda D A_k D^T$ & Ellipsoidal, equal volume \& orientation; variable shape \\
VVE & $\Sigma_k = \lambda_k D A_k D^T$ & Ellipsoidal, variable volume \& shape; equal orientation \\
EEV & $\Sigma_k = \lambda D_k A D_k^T$ & Ellipsoidal, equal volume \& shape; variable orientation \\
VEV & $\Sigma_k = \lambda_k D_k A D_k^T$ & Ellipsoidal, variable volume; equal shape; variable orientation \\
EVV & $\Sigma_k = \lambda D_k A_k D_k^T$ & Ellipsoidal, equal volume; variable shape \& orientation \\
VVV & $\Sigma_k = \lambda_k D_k A_k D_k^T$ & Ellipsoidal, variable volume, shape \& orientation \\
\bottomrule
\end{tabular}
\caption{The 14 Gaussian Parsimonious Clustering Models (GPCM) with covariance structures.}
\end{table}

\subsection{The $\alpha$--GPCM}
If we assume that the $\alpha$--transformed data (for any value of $\alpha \in \left[-1,1\right]$) follow a multivariate normal distribution, then a way to choose the value of $\alpha$ is via maximum likelihood estimation. The derivation of the $\alpha$ multivariate normal distribution is carried out by first assuming that $\bm{y}$ is multivariate normally distributed with parameters $\bm{\mu}_\alpha$ and $\bm{\Sigma}_\alpha$. If $\bm{y}\sim N_{p-1}\left(\bm{\mu}_\alpha,\bm{\Sigma}_\alpha \right )$, then the density of $\bm{y}$ is given by 
\begin{equation}
\label{MVN}
f_{\bm{y}}(\bm{y}) = \frac{1}{|2\pi \bm{\Sigma}_\alpha|^{1/2}}\exp{\left[-\frac{1}{2}\left(\bm{y}-\bm{\mu}_{\alpha}\right)^\top\bm{\Sigma}^{-1}_{\alpha}\left(\bm{y}-\bm{\mu}_{\alpha}\right)\right]}\left|J_{\alpha} \right|,
\end{equation}
and we can apply standard statistical techniques for finding the distribution of a transformation of a random vector to derive the distribution of $\mathbf{x}$. Recognise that in our case we require the  Jacobian of the $\alpha$--transformation, which is \citep{tsagris2020folded}
\begin{eqnarray}  \label{jac}
\left|J_{\alpha} \right|
=D^{D-1+\frac{1}{2}}\prod_{i=1}^p\frac{x_i^{\alpha-1}}{\sum_{j=1}^p x_j^{\alpha}}.
\end{eqnarray}

Application of the GPCM after transforming the compositional data by employing the $\alpha$--transformation yields the $\alpha$--GPCM. A possible drawback of this methodology is that zero values may still cause a problem since the Jacobian determinant (\ref{jac}) includes the product of the observed values and hence its logarithm would become infinity. Unlike the $\alpha$--$K$--means, the $\alpha$--transformation is not applicable, even for positive values of $\alpha$. The problem can be mitigated if one ignores the choice of $\alpha$ and chooses (a strictly positive) a value, say $\alpha=1$. In this case the BIC may still be used to decide the optimal model among the 14 GPCMs.  
 
\section{Alternative approaches}  \label{sec:alt}
Let us now list some alternative approaches that either utilize different power transformations and/or different mixture models.
\subsection{Alternative transformations}
Below is a list of power transformations that can replace the $\alpha$--transformation (\ref{alef}).

\begin{enumerate}
\item \cite{ait2003} defined the Box-Cox transformation applied to ratios of components
\begin{eqnarray*} 
f\left(\bm{x},\theta\right)=
\left\lbrace
\begin{array}{cc} 
\frac{\left(\frac{x_i}{x_p}\right)^{\lambda}-1}{\lambda} & \text{if $\lambda \neq 0$} \\
\log{\frac{x_i}{x_p}} &  \text{if $\lambda=0$}
\end{array} \right\rbrace,
\end{eqnarray*}
where the $x_d$ stands for the last component but since the ordering of the components is arbitrary, any component can play the role of the common divisor. 

\item A not so popular power transformation available in the literature is the folded-power transformation\footnote{This transformation has not been studied by any researcher in the context of compositional data (to the best of our knowledge).} \citep{atkinson1985}, extended to compositional data
\begin{eqnarray*} 
f\left(\bm{x},\lambda\right)=
\left\lbrace
\begin{array}{cc} 
\frac{x_i^\lambda -x_p^\lambda}{\lambda}, & \text{if} \ \ \lambda \neq 0 \\
\log{\frac{x_i}{x_p}},  & \text{if $\lambda=0$}
\end{array} \right\rbrace.
\end{eqnarray*}

\item \cite{greenacre2008, greenacre2010} applied the Box-Cox transformation to each of the components
\begin{eqnarray*}
f\left(\bm{x},\lambda\right)=
\left\lbrace
\begin{array}{cc} 
\frac{1}{\lambda}\left(x_i^{\lambda}-1\right) & \text{if $\lambda \neq 0$} \\
\log{x_i}                                     & \text{if $\lambda=0$}  
\end{array} \right\rbrace.
\end{eqnarray*}

\item \cite{scrucca2019} proposed the range-power transformation, in the context of model based clustering, for lower and upper bounded data
\begin{eqnarray*}
f\left(\bm{x},\lambda\right)=
\left\lbrace
\begin{array}{cc}
\frac{\left(\frac{x_i-l_i}{l_i-x_i}\right)^{\lambda}-1}{\lambda} & \text{if} \ \ \lambda \neq 0 \\
\log{\left(\frac{x_i-l_i}{l_i-x_i}\right)} & \text{if $\lambda = 0$}
\end{array}
\right\rbrace_{i=1,\ldots,D},
\end{eqnarray*}
where $l_i$ and $u_i$ denote the lower and upper bounds, respectively, of the $i$-th variable. In our case, the boundaries of the simple are given by 0 and 1 and hence $u_i=0$ and $l_i=1$ for $i=1,\ldots,p$. 

\item \cite{clarotto2022} defined the isometric $\alpha$--transformation
\begin{eqnarray*} 
f\left(\bm{x},\alpha\right)=
\left\lbrace
\begin{array}{cc}
\frac{1}{\alpha}\left(x_i^{\alpha} - \frac{\sum_{j=1}^px_{ij}^\alpha}{p}\right) & \text{if} \ \ \alpha \neq 0 \\
\log{x_i} - \frac{\sum_{j=1}^p\log{x_{ij}}}{p}  & \text{if $\alpha = 0$}
\end{array}
\right\rbrace.
\end{eqnarray*}

\item Finally, a power transformation that is parameter-free is the square root transformation \citep{stephens1982,scealy2011a}, applied to each component
\begin{eqnarray*}
f\left(\bm{x}\right)= \sqrt{x_i}.
\end{eqnarray*}
\end{enumerate}

\subsection{Alternative $K$--means algorithms and mixture models}
If one chooses the square root transformation, they may use the spherical $K$--means \citep{maitra2010,hornik2012}, mixtures of von Mises--Fisher distributions \citep{hornik2014,rossi2022} or mixtures of Spherical Cauchy and mixtures of Poisson kernel-based distributions \citep{tsagris2025,sablica2025}. Another approach is to use different mixture models, such as mixtures of $t$ distributions \citep{andrews2012}, mixtures of generalized hyperbolic distributions \citep{browne2015, wei2019} and mixtures of skew-$t$ distributions \citep{wei2019}. Note that in all these cases the 14 different parametrizations of the covariance matrix are still applicable. Evidently, these mixture models can be applied using the $\alpha$--transformation (\ref{alef}) or the aforementioned first 5 power transformations. 

The classical $K$--means may be robustified by using the trimmed $K$--means algorithm \citep{garcia2008} or the $K$--medoids \citep{rousseuw1987}. For the second case, appropriate CVIs include the silhouette index and the shadow values \citep{leisch2010}. Alternatively, one could use the Jensen-Shannon divergence metric or the Manhattan distance after the $\alpha$--transformation \citep{tsagris2014}. In a similar fashion, the GPCMs may be substituted by robust GPCMs \citep{punzo2018}.

When it comes to high-dimensional compositional data, the implementation of the GPCMs in the \textit{R} package \texttt{mixture} \citep{mixture2025} supports this case, but for the $K$--means, alternative options, e.g. regularisation \citep{raymaekers2022}, must be employed. Projecting the data onto lower dimensions is a strategy that works for both approaches \citep{boutsidis2010,berge2012}.

Another challenging example is large scale data or even big data. In these setting, mini-batch $K$--means \citep{sculley2010} should replace the classical $K$--means, with the trade-off being being between the optimal solution and the computational cost. Mini-batch variants yield sub-optimal solutions, yet magnitudes of order faster \citep{sculley2010}. 

\section{Simulation studies}  \label{sec:sim}
We generated compositional data from DMMs with $K=3, 4, 5, 6$ clusters, $p=3, 5, 10$ number of components, and various sample sizes, $n=300, 500, 1000, 2000, 50000, 10000$. For each triplet of number of clusters, dimensionality, and sample size we performed the $\alpha$--$K$--means and the $\alpha$--GPCM and tested the capability of each method using the absolute distance between the estimated number of cluster and the true number of clusters. In all cases, the number of clusters considered was 2 to 10. We further computed the number of times the value of $\alpha=0$, corresponding to the LRA approach, was selected by either CVI during $K$--means or by the BIC during the GPCMs. We repeated these steps 200 times and computed the average distance for each method and the average number of times a zero value of $\alpha$ was selected. 

Computational cost was another direction of comparison between the $K$--means and the GPCMs. We performed an independent study, based on 10 repetitions and report the average duration of either method using a single value of $\alpha$. For the $K$--means this includes the duration of the algorithm and the duration required to compute all 33 CVIs, whereas for the model-based clustering this includes the duration of fitting all 14 GPCMs. 

All computations took place in a desktop equipped with Intel Core i9-14900K at 3.2Ghz, 128GB RAM and Windows 11 Pro Installed using \textit{R} version 4.4.2. For the purpose of applicability and ease of reproducibility we have created an \textit{R} package that is available on github.   

\begin{table}[ht]
\centering
\caption{Results of the $\alpha$--$K$--means when $p=3$. Absolute distance between the estimated and the true number of clusters for the minimum valued CVIs.}
\label{minp3}
\begin{tabular}{l|rrrrrrrrr}
\toprule
 & \multicolumn{9}{c}{K=3} \\ \midrule 
Sample size & BRI & DBI & DRI & LDRI & LSSI & MRI & RTI & SSI & XBI \\  \midrule
n=300 & 7.00 & 0.00 & 1.00 & 1.00 & 1.00 & 7.00 & 0.00 & 7.00 & 0.04 \\ 
n=500 & 7.00 & 0.00 & 1.00 & 1.00 & 1.00 & 6.99 & 0.00 & 7.00 & 0.15 \\ 
n=1,000 & 7.00 & 0.00 & 1.00 & 1.00 & 1.00 & 7.00 & 0.00 & 7.00 & 0.06 \\ 
n=2,000 & 7.00 & 0.00 & 1.00 & 1.00 & 1.00 & 7.00 & 0.00 & 7.00 & 0.08 \\ 
n=5,000 & 7.00 & 0.00 & 1.00 & 1.00 & 1.00 & 7.00 & 0.00 & 7.00 & 0.95 \\ 
n=10,000 & 7.00 & 0.00 & 1.00 & 1.00 & 1.00 & 7.00 & 0.00 & 7.00 & 0.31 \\  \midrule
 & \multicolumn{9}{c}{K=4} \\ \midrule
Sample size & BRI & DBI & DRI & LDRI & LSSI & MRI & RTI & SSI & XBI \\  \midrule
n=300 & 6.00 & 0.00 & 2.00 & 2.00 & 2.00 & 6.00 & 0.00 & 6.00 & 1.60 \\ 
n=500 & 6.00 & 0.00 & 2.00 & 2.00 & 2.00 & 6.00 & 0.00 & 6.00 & 1.66 \\ 
n=1,000 & 6.00 & 0.00 & 2.00 & 2.00 & 2.00 & 5.99 & 0.00 & 6.00 & 1.88 \\ 
n=2,000 & 6.00 & 0.00 & 2.00 & 2.00 & 2.00 & 6.00 & 0.00 & 6.00 & 1.96 \\ 
n=5,000 & 6.00 & 0.00 & 2.00 & 2.00 & 2.00 & 6.00 & 0.00 & 6.00 & 1.98 \\ 
n=10,000 & 6.00 & 0.00 & 2.00 & 2.00 & 2.00 & 6.00 & 0.00 & 6.00 & 1.98 \\  \midrule
 & \multicolumn{9}{c}{K=5} \\ \midrule
Sample size & BRI & DBI & DRI & LDRI & LSSI & MRI & RTI & SSI & XBI \\  \midrule
n=300 & 5.00 & 0.00 & 3.00 & 3.00 & 3.00 & 4.99 & 0.02 & 4.99 & 2.19 \\ 
n=500 & 5.00 & 0.00 & 3.00 & 3.00 & 3.00 & 5.00 & 0.00 & 5.00 & 2.52 \\ 
n=1,000 & 4.99 & 0.00 & 3.00 & 3.00 & 3.00 & 5.00 & 0.00 & 4.99 & 2.83 \\ 
n=2,000 & 5.00 & 0.00 & 3.00 & 3.00 & 3.00 & 5.00 & 0.00 & 4.96 & 2.82 \\ 
n=5,000 & 5.00 & 0.00 & 3.00 & 3.00 & 3.00 & 5.00 & 0.00 & 4.99 & 2.94 \\ 
n=10,000 & 5.00 & 0.00 & 3.00 & 3.00 & 3.00 & 5.00 & 0.00 & 4.97 & 2.97 \\  \midrule
 & \multicolumn{9}{c}{K=6} \\ \midrule
Sample size & BRI & DBI & DRI & LDRI & LSSI & MRI & RTI & SSI & XBI \\  \midrule
n=300 & 4.00 & 0.00 & 4.00 & 4.00 & 4.00 & 3.99 & 0.02 & 4.00 & 2.65 \\ 
n=500 & 4.00 & 0.00 & 4.00 & 4.00 & 4.00 & 4.00 & 0.00 & 4.00 & 3.42 \\ 
n=1,000 & 4.00 & 0.00 & 4.00 & 4.00 & 4.00 & 4.00 & 0.00 & 4.00 & 3.15 \\ 
n=2,000 & 4.00 & 0.00 & 4.00 & 4.00 & 4.00 & 3.99 & 0.00 & 4.00 & 2.67 \\ 
n=5,000 & 4.00 & 0.00 & 4.00 & 4.00 & 4.00 & 3.99 & 0.00 & 4.00 & 3.32 \\ 
n=10,000 & 4.00 & 0.00 & 4.00 & 4.00 & 4.00 & 3.99 & 0.00 & 4.00 & 0.30 \\ 
\bottomrule
\end{tabular}
\end{table}

\begin{sidewaystable}
\centering
\caption{Results of the $\alpha$--$K$--means when $p=3$. Absolute distance between the estimated and the true number of clusters for the maximum valued CVIs.}
\label{maxp3}
\setlength{\tabcolsep}{2.5pt} 
\begin{small}
\begin{tabular}{l|rrrrrrrrrrrrrrrrrrrrrrrrr}
  \toprule
 & \multicolumn{24}{c}{K=3} \\ \midrule 
Sample size & BHI & CHI & DI & GDI11 & GDI12 & GDI13 & GDI21 & GDI22 & GDI23 & GDI31 & GDI32 & GDI33 & GDI41 & GDI42 & GDI43 & GDI51 & GDI52 & GDI53 & KWI & PBMI & PBI & RLI & SI & TWBI \\ 
\midrule
n=300 & 1.00 & 2.61 & 0.08 & 0.08 & 0.04 & 0.02 & 0.38 & 0.29 & 0.26 & 0.00 & 0.00 & 0.00 & 0.00 & 0.00 & 0.00 & 1.62 & 0.37 & 0.26 & 1.00 & 0.99 & 7.00 & 1.00 & 0.00 & 7.00 \\ 
n=500 & 1.00 & 3.45 & 0.16 & 0.16 & 0.14 & 0.04 & 0.14 & 0.13 & 0.11 & 0.00 & 0.00 & 0.00 & 0.00 & 0.00 & 0.00 & 1.09 & 0.05 & 0.04 & 1.00 & 0.49 & 7.00 & 1.00 & 0.00 & 7.00 \\ 
n=1,000 & 1.00 & 1.53 & 0.06 & 0.06 & 0.04 & 0.04 & 0.76 & 0.84 & 0.87 & 0.00 & 0.00 & 0.00 & 0.00 & 0.00 & 0.00 & 1.00 & 0.92 & 1.40 & 1.00 & 1.00 & 7.00 & 1.00 & 0.00 & 7.00 \\ 
n=2,000 & 1.00 & 1.82 & 0.41 & 0.41 & 0.08 & 0.06 & 0.94 & 0.90 & 0.87 & 0.00 & 0.00 & 0.00 & 0.00 & 0.00 & 0.00 & 1.02 & 0.00 & 0.00 & 1.00 & 1.00 & 7.00 & 1.00 & 0.00 & 7.00 \\ 
n=5,000 & 1.00 & 1.97 & 0.97 & 0.97 & 0.95 & 0.94 & 0.08 & 0.05 & 0.05 & 0.00 & 0.00 & 0.00 & 0.00 & 0.00 & 0.00 & 1.00 & 0.00 & 0.00 & 1.00 & 1.00 & 7.00 & 1.00 & 0.00 & 7.00 \\ 
n=10,000 & 1.00 & 1.57 & 0.88 & 0.88 & 0.27 & 0.27 & 0.28 & 0.09 & 0.09 & 0.00 & 0.00 & 0.00 & 0.00 & 0.00 & 0.00 & 1.00 & 0.00 & 0.00 & 1.00 & 1.00 & 7.00 & 1.00 & 0.00 & 7.00 \\
\midrule
 & \multicolumn{24}{c}{K=4} \\ \midrule
Sample size & BHI & CHI & DI & GDI11 & GDI12 & GDI13 & GDI21 & GDI22 & GDI23 & GDI31 & GDI32 & GDI33 & GDI41 & GDI42 & GDI43 & GDI51 & GDI52 & GDI53 & KWI & PBMI & PBI & RLI & SI & TWBI \\ \midrule
n=300 & 2.00 & 2.49 & 1.76 & 1.76 & 1.72 & 1.62 & 0.95 & 0.04 & 0.01 & 0.36 & 0.00 & 0.00 & 0.20 & 0.00 & 0.00 & 2.00 & 2.00 & 2.00 & 1.11 & 0.00 & 6.00 & 1.00 & 0.00 & 6.00 \\ 
n=500 & 2.00 & 4.83 & 1.78 & 1.78 & 1.76 & 1.72 & 0.98 & 0.03 & 0.06 & 0.92 & 0.00 & 0.00 & 0.88 & 0.00 & 0.00 & 2.00 & 2.00 & 2.00 & 1.02 & 0.00 & 6.00 & 1.00 & 0.00 & 6.00 \\ 
n=1,000 & 2.00 & 1.43 & 1.96 & 1.96 & 1.94 & 1.90 & 0.97 & 0.02 & 0.01 & 0.44 & 0.00 & 0.00 & 0.25 & 0.00 & 0.00 & 2.00 & 2.00 & 2.00 & 1.01 & 0.00 & 5.96 & 1.00 & 0.00 & 5.99 \\ 
n=2000 & 2.00 & 0.18 & 1.98 & 1.98 & 1.96 & 1.96 & 1.04 & 0.02 & 0.02 & 0.97 & 0.00 & 0.00 & 0.95 & 0.00 & 0.00 & 2.00 & 2.00 & 2.00 & 1.02 & 0.00 & 5.96 & 1.00 & 0.00 & 6.00 \\ 
n=5,000 & 2.00 & 0.00 & 2.00 & 2.00 & 1.98 & 1.98 & 1.00 & 0.03 & 0.03 & 0.48 & 0.00 & 0.00 & 0.29 & 0.00 & 0.00 & 2.00 & 2.00 & 2.00 & 1.03 & 0.00 & 6.00 & 1.00 & 0.00 & 6.00 \\ 
n=10,000 & 2.00 & 0.00 & 2.00 & 2.00 & 1.98 & 1.98 & 1.00 & 0.03 & 0.02 & 0.99 & 0.00 & 0.00 & 0.97 & 0.00 & 0.00 & 2.00 & 2.00 & 2.00 & 1.00 & 0.00 & 6.00 & 1.00 & 0.00 & 6.00 \\ 
\midrule
 & \multicolumn{24}{c}{K=5} \\ \midrule
Sample size & BHI & CHI & DI & GDI11 & GDI12 & GDI13 & GDI21 & GDI22 & GDI23 & GDI31 & GDI32 & GDI33 & GDI41 & GDI42 & GDI43 & GDI51 & GDI52 & GDI53 & KWI & PBMI & PBI & RLI & SI & TWBI \\  \midrule
n=300 & 3.00 & 4.12 & 2.74 & 2.74 & 2.71 & 2.54 & 1.48 & 0.50 & 0.40 & 0.35 & 0.03 & 0.00 & 0.33 & 0.03 & 0.00 & 2.78 & 2.99 & 3.00 & 3.00 & 0.66 & 5.00 & 3.00 & 0.00 & 4.99 \\ 
n=500.2 & 3.00 & 3.97 & 2.87 & 2.87 & 2.74 & 2.69 & 1.96 & 2.31 & 1.99 & 0.40 & 0.00 & 0.00 & 0.42 & 0.00 & 0.00 & 2.44 & 2.88 & 3.00 & 3.00 & 1.37 & 5.00 & 3.00 & 0.00 & 5.00 \\ 
n=1,000 & 3.00 & 4.32 & 2.96 & 2.96 & 2.96 & 2.93 & 2.03 & 1.24 & 0.76 & 0.81 & 0.00 & 0.00 & 0.46 & 0.00 & 0.00 & 2.55 & 2.99 & 3.00 & 3.00 & 0.11 & 5.00 & 3.00 & 0.00 & 5.00 \\ 
n=2,000 & 3.00 & 3.15 & 2.95 & 2.95 & 2.88 & 2.84 & 1.85 & 0.85 & 0.48 & 0.31 & 0.00 & 0.00 & 0.33 & 0.00 & 0.00 & 2.29 & 3.00 & 3.00 & 3.00 & 0.03 & 5.00 & 3.00 & 0.00 & 5.00 \\ 
n=5,000 & 3.00 & 3.04 & 2.99 & 2.99 & 2.99 & 2.93 & 2.07 & 0.42 & 0.24 & 1.67 & 0.00 & 0.00 & 1.60 & 0.00 & 0.00 & 2.23 & 3.00 & 3.00 & 3.00 & 0.00 & 5.00 & 3.00 & 0.00 & 5.00 \\ 
n=10,000 & 3.00 & 3.04 & 3.00 & 3.00 & 2.97 & 2.97 & 2.90 & 0.96 & 0.88 & 1.06 & 0.00 & 0.00 & 1.05 & 0.00 & 0.00 & 2.93 & 3.00 & 3.00 & 3.00 & 0.00 & 5.00 & 3.00 & 0.00 & 5.00 \\
\midrule
 & \multicolumn{24}{c}{K=6} \\ \midrule
Sample size & BHI & CHI & DI & GDI11 & GDI12 & GDI13 & GDI21 & GDI22 & GDI23 & GDI31 & GDI32 & GDI33 & GDI41 & GDI42 & GDI43 & GDI51 & GDI52 & GDI53 & KWI & PBMI & PBI & RLI & SI & TWBI \\  \midrule
n=300  & 4.00 & 3.72 & 3.26 & 3.26 & 3.06 & 3.14 & 2.34 & 0.78 & 0.52 & 1.95 & 0.04 & 0.00 & 1.86 & 0.02 & 0.00 & 3.41 & 3.60 & 3.84 & 3.99 & 0.20 & 4.00 & 3.00 & 0.00 & 4.00 \\ 
n=500 & 4.00 & 3.72 & 3.69 & 3.69 & 3.57 & 3.55 & 2.15 & 1.18 & 0.18 & 1.51 & 0.00 & 0.00 & 1.41 & 0.00 & 0.00 & 3.73 & 3.93 & 4.00 & 4.00 & 0.08 & 4.00 & 3.00 & 0.00 & 4.00 \\ 
n=1,000 & 4.00 & 3.72 & 3.33 & 3.33 & 3.33 & 3.22 & 2.79 & 0.38 & 0.07 & 2.72 & 0.00 & 0.00 & 2.59 & 0.00 & 0.00 & 3.48 & 3.83 & 3.98 & 4.00 & 0.00 & 4.00 & 3.00 & 0.00 & 4.00 \\ 
n=2,000 & 4.00 & 3.65 & 2.81 & 2.81 & 2.65 & 2.61 & 2.46 & 0.31 & 0.13 & 1.77 & 0.00 & 0.00 & 1.34 & 0.00 & 0.00 & 3.54 & 3.89 & 4.00 & 4.00 & 0.03 & 4.00 & 3.00 & 0.00 & 4.00 \\ 
n=5,000 & 4.00 & 3.88 & 3.40 & 3.40 & 3.31 & 3.31 & 2.83 & 0.17 & 0.03 & 2.24 & 0.00 & 0.00 & 2.18 & 0.00 & 0.00 & 3.30 & 3.79 & 3.98 & 4.00 & 0.00 & 4.00 & 3.00 & 0.00 & 4.00 \\ 
n=10,000 & 4.00 & 3.72 & 0.41 & 0.41 & 0.27 & 0.27 & 2.52 & 1.09 & 0.14 & 2.97 & 0.00 & 0.00 & 2.95 & 0.00 & 0.00 & 3.91 & 3.92 & 4.00 & 4.00 & 0.00 & 4.00 & 3.00 & 0.00 & 4.00 \\ 
\bottomrule
\end{tabular}
\end{small}
\end{sidewaystable}

\begin{table}[ht]
\centering
\caption{Results of the $\alpha$--$K$--means when $p=5$. Absolute distance between the estimated and the true number of clusters for the minimum valued CVIs.}
\label{minp5}
\begin{tabular}{l|rrrrrrrrr}
\toprule
 & \multicolumn{9}{c}{K=3} \\ \midrule 
Sample size & BRI & DBI & DRI & LDRI & LSSI & MRI & RTI & SSI & XBI \\  \midrule
n=300 & 7.00 & 0.00 & 1.00 & 1.00 & 1.00 & 6.98 & 0.00 & 6.95 & 0.00 \\ 
n=500 & 7.00 & 0.00 & 1.00 & 1.00 & 1.00 & 7.00 & 0.00 & 7.00 & 0.03 \\ 
n=1000 & 7.00 & 0.00 & 1.00 & 1.00 & 1.00 & 7.00 & 0.00 & 7.00 & 0.00 \\ 
n=2000 & 7.00 & 0.00 & 1.00 & 1.00 & 1.00 & 6.99 & 0.00 & 7.00 & 0.00 \\ 
n=5000 & 7.00 & 0.00 & 1.00 & 1.00 & 1.00 & 6.99 & 0.00 & 7.00 & 0.00 \\ 
n=10000 & 7.00 & 0.00 & 1.00 & 1.00 & 1.00 & 7.00 & 0.00 & 7.00 & 0.01 \\ 
\midrule
& \multicolumn{9}{c}{K=4} \\ \midrule
Sample size & BRI & DBI & DRI & LDRI & LSSI & MRI & RTI & SSI & XBI \\  \midrule
n=300 & 6.00 & 0.00 & 2.00 & 2.00 & 2.00 & 5.70 & 0.00 & 5.99 & 0.00 \\ 
n=500 & 6.00 & 0.00 & 2.00 & 2.00 & 2.00 & 5.84 & 0.00 & 6.00 & 0.00 \\ 
n=1000 & 6.00 & 0.00 & 2.00 & 2.00 & 2.00 & 5.91 & 0.00 & 6.00 & 0.00 \\ 
n=2000 & 6.00 & 0.00 & 2.00 & 2.00 & 2.00 & 5.98 & 0.00 & 6.00 & 0.00 \\ 
n=5000 & 6.00 & 0.00 & 2.00 & 2.00 & 2.00 & 5.90 & 0.00 & 6.00 & 0.00 \\ 
n=10000 & 6.00 & 0.00 & 2.00 & 2.00 & 2.00 & 5.96 & 0.00 & 6.00 & 0.00 \\ 
\midrule
& \multicolumn{9}{c}{K=5} \\ \midrule
Sample size & BRI & DBI & DRI & LDRI & LSSI & MRI & RTI & SSI & XBI \\  \midrule
n=300 & 5.00 & 0.00 & 3.00 & 3.00 & 3.00 & 4.97 & 0.00 & 4.97 & 0.12 \\ 
n=500 & 5.00 & 0.00 & 3.00 & 3.00 & 3.00 & 4.98 & 0.00 & 4.99 & 0.56 \\ 
n=1000 & 5.00 & 0.00 & 3.00 & 3.00 & 3.00 & 4.97 & 0.00 & 4.99 & 1.40 \\ 
n=2000 & 5.00 & 0.00 & 3.00 & 3.00 & 3.00 & 4.94 & 0.00 & 4.97 & 0.23 \\ 
n=5000 & 5.00 & 0.00 & 3.00 & 3.00 & 3.00 & 4.88 & 0.00 & 5.00 & 0.28 \\ 
n=10000 & 5.00 & 0.00 & 3.00 & 3.00 & 3.00 & 4.99 & 0.00 & 4.99 & 1.54 \\ 
\midrule
& \multicolumn{9}{c}{K=6} \\ \midrule
Sample size & BRI & DBI & DRI & LDRI & LSSI & MRI & RTI & SSI & XBI \\  \midrule
n=300 & 4.00 & 0.72 & 4.00 & 4.00 & 4.00 & 3.59 & 0.00 & 3.94 & 0.58 \\ 
n=500 & 4.00 & 0.20 & 4.00 & 4.00 & 4.00 & 3.82 & 0.00 & 3.96 & 0.52 \\ 
n=1000 & 4.00 & 0.36 & 4.00 & 4.00 & 4.00 & 3.81 & 0.00 & 3.95 & 2.67 \\ 
n=2000 & 4.00 & 0.12 & 4.00 & 4.00 & 4.00 & 3.83 & 0.00 & 3.90 & 3.64 \\ 
n=5000 & 4.00 & 0.00 & 4.00 & 4.00 & 4.00 & 3.65 & 0.00 & 3.80 & 3.92 \\ 
n=10000& 4.00 & 0.00 & 4.00 & 4.00 & 4.00 & 3.83 & 0.00 & 3.94 & 4.00 \\ 
\bottomrule
\end{tabular}
\end{table}

\begin{sidewaystable}
\centering
\caption{Results of the $\alpha$--$K$--means when $p=5$. Absolute distance between the estimated and the true number of clusters for the maximum valued CVIs.}
\label{maxp5}
\setlength{\tabcolsep}{2.5pt} 
\begin{small}
\begin{tabular}{l|rrrrrrrrrrrrrrrrrrrrrrrrr}
  \toprule
 & \multicolumn{24}{c}{K=3} \\ \midrule 
Sample size & BHI & CHI & DI & GDI11 & GDI12 & GDI13 & GDI21 & GDI22 & GDI23 & GDI31 & GDI32 & GDI33 & GDI41 & GDI42 & GDI43 & GDI51 & GDI52 & GDI53 & KWI & PBMI & PBI & RLI & SI & TWBI \\ 
\midrule
n=300 & 1.00 & 0.00 & 0.00 & 0.00 & 0.00 & 0.00 & 0.03 & 0.88 & 0.89 & 0.00 & 0.00 & 0.00 & 0.00 & 0.00 & 0.00 & 4.14 & 1.16 & 1.22 & 1.00 & 0.00 & 6.99 & 0.00 & 0.00 & 6.95 \\ 
n=500 & 1.00 & 0.00 & 0.26 & 0.26 & 0.17 & 0.01 & 0.11 & 0.17 & 0.17 & 0.00 & 0.00 & 0.00 & 0.00 & 0.00 & 0.00 & 2.90 & 1.16 & 1.26 & 1.00 & 0.00 & 7.00 & 0.00 & 0.00 & 6.99 \\ 
n=1000 & 1.00 & 0.00 & 0.00 & 0.00 & 0.00 & 0.00 & 0.04 & 0.94 & 0.94 & 0.00 & 0.00 & 0.00 & 0.00 & 0.00 & 0.00 & 4.31 & 1.03 & 1.04 & 1.00 & 0.00 & 7.00 & 0.00 & 0.00 & 7.00 \\ 
n=2000 & 1.00 & 0.00 & 0.01 & 0.01 & 0.00 & 0.00 & 0.08 & 0.99 & 0.99 & 0.00 & 0.00 & 0.00 & 0.00 & 0.00 & 0.00 & 1.43 & 1.00 & 1.00 & 1.00 & 0.00 & 7.00 & 0.00 & 0.00 & 7.00 \\ 
n=5000 & 1.00 & 0.00 & 0.00 & 0.00 & 0.00 & 0.00 & 0.52 & 1.00 & 1.00 & 0.00 & 0.00 & 0.00 & 0.00 & 0.00 & 0.00 & 1.28 & 1.00 & 1.00 & 1.00 & 0.00 & 7.00 & 0.00 & 0.00 & 6.99 \\ 
n=10000 & 1.00 & 0.00 & 0.01 & 0.01 & 0.01 & 0.01 & 0.09 & 1.00 & 1.00 & 0.00 & 0.00 & 0.00 & 0.00 & 0.00 & 0.00 & 1.18 & 1.00 & 1.00 & 1.00 & 0.00 & 7.00 & 0.00 & 0.00 & 7.00 \\
  \midrule
 & \multicolumn{24}{c}{K=4} \\ \midrule
Sample size & BHI & CHI & DI & GDI11 & GDI12 & GDI13 & GDI21 & GDI22 & GDI23 & GDI31 & GDI32 & GDI33 & GDI41 & GDI42 & GDI43 & GDI51 & GDI52 & GDI53 & KWI & PBMI & PBI & RLI & SI & TWBI \\  \midrule
n=300 & 2.00 & 0.00 & 0.03 & 0.03 & 0.00 & 0.00 & 0.00 & 0.08 & 0.04 & 0.00 & 0.00 & 0.01 & 0.01 & 0.01 & 0.01 & 2.00 & 2.00 & 2.00 & 2.00 & 0.00 & 6.00 & 1.00 & 0.07 & 5.88 \\ 
n=500 & 2.00 & 0.00 & 0.00 & 0.00 & 0.00 & 0.00 & 0.03 & 0.49 & 0.39 & 0.01 & 0.00 & 0.00 & 0.01 & 0.00 & 0.00 & 2.00 & 2.00 & 2.00 & 2.00 & 0.00 & 6.00 & 1.00 & 0.00 & 5.85 \\ 
n=1000 & 2.00 & 0.00 & 0.00 & 0.00 & 0.00 & 0.00 & 0.02 & 0.65 & 0.07 & 0.01 & 0.00 & 0.00 & 0.02 & 0.00 & 0.00 & 2.00 & 2.00 & 2.00 & 2.00 & 0.00 & 6.00 & 1.00 & 0.00 & 5.98 \\ 
n=2000 & 2.00 & 0.00 & 0.02 & 0.02 & 0.00 & 0.00 & 0.03 & 0.85 & 0.75 & 0.00 & 0.00 & 0.00 & 0.03 & 0.00 & 0.00 & 2.00 & 2.00 & 2.00 & 2.00 & 0.00 & 6.00 & 1.00 & 0.00 & 5.97 \\ 
n=5000 & 2.00 & 0.00 & 0.00 & 0.00 & 0.00 & 0.00 & 0.00 & 0.91 & 0.80 & 0.00 & 0.00 & 0.00 & 0.02 & 0.00 & 0.00 & 2.00 & 2.00 & 2.00 & 2.00 & 0.00 & 6.00 & 1.00 & 0.00 & 5.98 \\ 
n=10000 & 2.00 & 0.00 & 0.00 & 0.00 & 0.00 & 0.00 & 0.01 & 0.66 & 0.16 & 0.00 & 0.00 & 0.00 & 0.04 & 0.00 & 0.00 & 2.00 & 2.00 & 2.00 & 2.00 & 0.00 & 6.00 & 1.00 & 0.00 & 6.00 \\
  \midrule
 & \multicolumn{24}{c}{K=5} \\ \midrule
Sample size & BHI & CHI & DI & GDI11 & GDI12 & GDI13 & GDI21 & GDI22 & GDI23 & GDI31 & GDI32 & GDI33 & GDI41 & GDI42 & GDI43 & GDI51 & GDI52 & GDI53 & KWI & PBMI & PBI & RLI & SI & TWBI \\  \midrule
n=300 & 3.00 & 0.00 & 0.09 & 0.09 & 0.15 & 0.11 & 0.64 & 0.96 & 0.42 & 0.00 & 0.06 & 0.00 & 0.00 & 0.06 & 0.00 & 3.00 & 3.00 & 2.99 & 3.00 & 0.15 & 4.94 & 2.96 & 0.00 & 4.73 \\ 
n=500 & 3.00 & 0.00 & 1.38 & 1.38 & 0.65 & 0.63 & 2.62 & 2.55 & 1.30 & 0.52 & 0.00 & 0.00 & 0.37 & 0.00 & 0.00 & 2.96 & 3.00 & 2.99 & 3.00 & 0.00 & 4.96 & 3.00 & 0.00 & 4.94 \\ 
n=1000 & 3.00 & 0.00 & 1.48 & 1.48 & 1.46 & 1.44 & 2.66 & 2.82 & 2.59 & 0.42 & 0.00 & 0.00 & 0.24 & 0.00 & 0.00 & 3.00 & 3.00 & 3.00 & 3.00 & 0.02 & 4.99 & 3.00 & 0.00 & 4.70 \\ 
n=2000 & 3.00 & 0.00 & 0.40 & 0.40 & 0.41 & 0.29 & 1.32 & 2.79 & 2.44 & 0.06 & 0.00 & 0.00 & 0.06 & 0.00 & 0.00 & 3.00 & 3.00 & 3.00 & 3.00 & 0.00 & 4.98 & 3.00 & 0.00 & 4.84 \\ 
n=5000 & 3.00 & 0.00 & 0.38 & 0.38 & 0.38 & 0.34 & 2.83 & 1.48 & 1.19 & 0.26 & 0.00 & 0.00 & 0.20 & 0.00 & 0.00 & 3.00 & 3.00 & 3.00 & 3.00 & 0.00 & 5.00 & 3.00 & 0.00 & 4.99 \\ 
n=10000 & 3.00 & 0.00 & 1.64 & 1.64 & 1.76 & 1.70 & 2.74 & 2.85 & 2.68 & 0.29 & 0.00 & 0.00 & 0.23 & 0.00 & 0.00 & 3.00 & 3.00 & 3.00 & 3.00 & 0.00 & 5.00 & 3.00 & 0.00 & 4.99 \\ 
  \midrule
 & \multicolumn{24}{c}{K=6} \\ \midrule
Sample size & BHI & CHI & DI & GDI11 & GDI12 & GDI13 & GDI21 & GDI22 & GDI23 & GDI31 & GDI32 & GDI33 & GDI41 & GDI42 & GDI43 & GDI51 & GDI52 & GDI53 & KWI & PBMI & PBI & RLI & SI & TWBI \\  \midrule
n=300 & 4.00 & 0.00 & 0.84 & 0.84 & 1.88 & 1.54 & 3.60 & 4.00 & 4.00 & 0.11 & 1.88 & 0.56 & 0.06 & 2.08 & 0.56 & 4.00 & 4.00 & 4.00 & 4.00 & 0.00 & 4.00 & 3.95 & 1.56 & 3.96 \\ 
n=500 & 4.00 & 0.00 & 0.68 & 0.68 & 1.00 & 0.78 & 3.56 & 4.00 & 4.00 & 0.00 & 0.96 & 0.16 & 0.00 & 0.84 & 0.08 & 4.00 & 3.99 & 4.00 & 4.00 & 0.00 & 4.00 & 3.99 & 0.92 & 3.93 \\ 
n=1000 & 4.00 & 0.00 & 3.52 & 3.52 & 3.74 & 3.62 & 3.76 & 4.00 & 4.00 & 0.00 & 3.36 & 2.84 & 0.00 & 3.28 & 2.08 & 4.00 & 4.00 & 4.00 & 4.00 & 0.00 & 4.00 & 3.98 & 1.40 & 3.57 \\ 
n=2000 & 4.00 & 0.00 & 3.96 & 3.96 & 3.96 & 3.88 & 3.96 & 4.00 & 4.00 & 0.04 & 0.52 & 0.12 & 0.04 & 0.32 & 0.04 & 4.00 & 4.00 & 4.00 & 4.00 & 0.00 & 4.00 & 4.00 & 1.76 & 3.66 \\ 
n=5000 & 4.00 & 0.00 & 4.00 & 4.00 & 4.00 & 4.00 & 3.96 & 4.00 & 4.00 & 0.21 & 3.24 & 0.00 & 0.22 & 0.96 & 0.00 & 4.00 & 4.00 & 4.00 & 4.00 & 0.00 & 4.00 & 4.00 & 1.72 & 3.87 \\ 
n=10000 & 4.00 & 0.00 & 4.00 & 4.00 & 4.00 & 4.00 & 4.00 & 4.00 & 4.00 & 0.20 & 3.48 & 0.00 & 0.16 & 2.24 & 0.00 & 4.00 & 4.00 & 4.00 & 4.00 & 0.00 & 4.00 & 4.00 & 0.00 & 3.72 \\ 
\bottomrule
\end{tabular}
\end{small}
\end{sidewaystable}

\begin{table}[ht]
\centering
\caption{Results of the $\alpha$--$K$--means when $p=10$. Absolute distance between the estimated and the true number of clusters for the minimum valued CVIs.}
\label{minp10}
\begin{tabular}{l|rrrrrrrrr}
\toprule
 & \multicolumn{9}{c}{K=3} \\ \midrule 
Sample size & BRI & DBI & DRI & LDRI & LSSI & MRI & RTI & SSI & XBI \\  \midrule
n=300 & 7.00 & 0.00 & 1.00 & 1.00 & 1.00 & 1.34 & 0.00 & 7.00 & 0.00 \\ 
n=500 & 7.00 & 0.00 & 1.00 & 1.00 & 1.00 & 1.90 & 0.00 & 7.00 & 0.00 \\ 
n=1000 & 7.00 & 0.00 & 1.00 & 1.00 & 1.00 & 1.14 & 0.00 & 7.00 & 0.00 \\ 
n=2000 & 7.00 & 0.00 & 1.00 & 1.00 & 1.00 & 1.05 & 0.00 & 7.00 & 0.00 \\ 
n=5000 & 7.00 & 0.00 & 1.00 & 1.00 & 1.00 & 1.02 & 0.00 & 7.00 & 0.00 \\ 
n=10000 & 7.00 & 0.00 & 1.00 & 1.00 & 1.00 & 1.04 & 0.00 & 7.00 & 0.00 \\  \midrule
 & \multicolumn{9}{c}{K=4} \\ \midrule
Sample size & BRI & DBI & DRI & LDRI & LSSI & MRI & RTI & SSI & XBI \\  \midrule
n=300 & 6.00 & 0.00 & 2.00 & 2.00 & 2.00 & 4.84 & 0.00 & 5.88 & 0.00 \\ 
n=500 & 6.00 & 0.00 & 2.00 & 2.00 & 2.00 & 4.74 & 0.00 & 6.00 & 0.00 \\ 
n=1000 & 6.00 & 0.00 & 2.00 & 2.00 & 2.00 & 3.60 & 0.00 & 5.94 & 0.00 \\ 
n=2000 & 6.00 & 0.00 & 2.00 & 2.00 & 2.00 & 3.24 & 0.00 & 5.94 & 0.00 \\ 
n=5000 & 6.00 & 0.00 & 2.00 & 2.00 & 2.00 & 2.74 & 0.00 & 5.94 & 0.00 \\ 
n=10000 & 6.00 & 0.00 & 2.00 & 2.00 & 2.00 & 3.44 & 0.00 & 5.80 & 0.00 \\ \midrule
   & \multicolumn{9}{c}{K=5} \\ \midrule
Sample size & BRI & DBI & DRI & LDRI & LSSI & MRI & RTI & SSI & XBI \\  \midrule
n=300 & 4.58 & 3.00 & 3.00 & 3.00 & 3.00 & 3.80 & 2.40 & 4.38 & 2.52 \\ 
n=500 & 4.90 & 3.00 & 3.00 & 3.00 & 3.00 & 4.26 & 2.82 & 4.84 & 2.76 \\ 
n=1000 & 5.00 & 3.00 & 3.00 & 3.00 & 3.00 & 4.28 & 3.00 & 4.88 & 3.00 \\ 
n=2000 & 4.94 & 3.00 & 3.00 & 3.00 & 3.00 & 4.22 & 3.00 & 4.78 & 3.00 \\ 
n=5000 & 5.00 & 3.00 & 3.00 & 3.00 & 3.00 & 4.02 & 3.00 & 4.74 & 3.00 \\ 
n=10000 & 4.96 & 3.00 & 3.00 & 3.00 & 3.00 & 4.16 & 3.00 & 4.88 & 3.00 \\  \midrule
 & \multicolumn{9}{c}{K=6} \\ \midrule
Sample size & BRI & DBI & DRI & LDRI & LSSI & MRI & RTI & SSI & XBI \\  \midrule
n=300 & 3.96 & 2.24 & 4.00 & 4.00 & 4.00 & 3.06 & 0.00 & 2.52 & 0.02 \\ 
n=500 & 4.00 & 2.24 & 4.00 & 4.00 & 4.00 & 3.40 & 0.00 & 3.56 & 0.00 \\ 
n=1000 & 4.00 & 2.32 & 4.00 & 4.00 & 4.00 & 3.54 & 0.00 & 3.78 & 0.04 \\ 
n=2000 & 4.00 & 2.72 & 4.00 & 4.00 & 4.00 & 3.56 & 0.00 & 3.72 & 0.04 \\ 
n=5000 & 4.00 & 3.12 & 4.00 & 4.00 & 4.00 & 3.54 & 0.00 & 3.72 & 0.00 \\ 
n=10000 & 4.00 & 3.76 & 4.00 & 4.00 & 4.00 & 3.48 & 0.00 & 3.40 & 0.00 \\ 
\bottomrule
\end{tabular}
\end{table}

\begin{sidewaystable}
\centering
\caption{Results of the $\alpha$--$K$--means when $p=10$. Absolute distance between the estimated and the true number of clusters for the maximum valued CVIs.}
\label{maxp10}
\setlength{\tabcolsep}{2.5pt} 
\begin{small}
\begin{tabular}{l|rrrrrrrrrrrrrrrrrrrrrrrrr}
  \toprule
 & \multicolumn{24}{c}{K=3} \\ \midrule 
Sample size & BHI & CHI & DI & GDI11 & GDI12 & GDI13 & GDI21 & GDI22 & GDI23 & GDI31 & GDI32 & GDI33 & GDI41 & GDI42 & GDI43 & GDI51 & GDI52 & GDI53 & KWI & PBMI & PBI & RLI & SI & TWBI \\ 
\midrule
n=300 & 1.00 & 0.00 & 0.00 & 0.00 & 0.00 & 0.00 & 0.01 & 0.00 & 0.01 & 0.00 & 0.00 & 0.00 & 0.00 & 0.00 & 0.00 & 3.50 & 1.40 & 1.78 & 1.00 & 0.00 & 7.00 & 0.00 & 0.00 & 6.69 \\ 
n=500 & 1.00 & 0.00 & 0.00 & 0.00 & 0.00 & 0.00 & 0.00 & 0.02 & 0.03 & 0.00 & 0.00 & 0.00 & 0.00 & 0.00 & 0.00 & 3.65 & 1.69 & 2.72 & 1.00 & 0.00 & 7.00 & 0.00 & 0.00 & 6.71 \\ 
n=1000 & 1.00 & 0.00 & 0.00 & 0.00 & 0.00 & 0.00 & 0.00 & 0.46 & 0.54 & 0.00 & 0.00 & 0.00 & 0.00 & 0.00 & 0.00 & 2.08 & 1.37 & 1.45 & 1.00 & 0.00 & 7.00 & 0.00 & 0.00 & 6.89 \\ 
n=2000 & 1.00 & 0.00 & 0.00 & 0.00 & 0.00 & 0.00 & 0.00 & 0.64 & 0.69 & 0.00 & 0.00 & 0.00 & 0.00 & 0.00 & 0.00 & 2.03 & 1.67 & 1.75 & 1.00 & 0.00 & 7.00 & 0.00 & 0.00 & 6.81 \\ 
n=5000 & 1.00 & 0.00 & 0.00 & 0.00 & 0.00 & 0.00 & 0.00 & 0.99 & 0.96 & 0.00 & 0.00 & 0.00 & 0.00 & 0.00 & 0.00 & 1.07 & 1.98 & 1.82 & 1.00 & 0.00 & 7.00 & 0.00 & 0.00 & 6.87 \\ 
n=10000 & 1.00 & 0.00 & 0.00 & 0.00 & 0.00 & 0.00 & 0.00 & 0.92 & 0.93 & 0.00 & 0.00 & 0.00 & 0.00 & 0.00 & 0.00 & 1.02 & 1.69 & 1.75 & 1.00 & 0.00 & 7.00 & 0.00 & 0.00 & 6.87 \\ 
  \midrule
 & \multicolumn{24}{c}{K=4} \\ \midrule
Sample size & BHI & CHI & DI & GDI11 & GDI12 & GDI13 & GDI21 & GDI22 & GDI23 & GDI31 & GDI32 & GDI33 & GDI41 & GDI42 & GDI43 & GDI51 & GDI52 & GDI53 & KWI & PBMI & PBI & RLI & SI & TWBI \\  \midrule
n=300 & 2.00 & 0.00 & 0.00 & 0.00 & 0.00 & 0.00 & 0.04 & 0.20 & 0.12 & 0.00 & 0.00 & 0.00 & 0.00 & 0.00 & 0.00 & 2.06 & 2.00 & 2.00 & 2.00 & 0.00 & 5.98 & 1.34 & 0.00 & 5.62 \\ 
n=500 & 2.00 & 0.00 & 0.00 & 0.00 & 0.00 & 0.00 & 0.00 & 0.18 & 0.10 & 0.00 & 0.00 & 0.00 & 0.00 & 0.00 & 0.00 & 2.00 & 2.00 & 2.00 & 2.00 & 0.00 & 5.94 & 1.26 & 0.00 & 5.72 \\ 
n=1000 & 2.00 & 0.00 & 0.00 & 0.00 & 0.00 & 0.00 & 0.08 & 0.18 & 0.14 & 0.00 & 0.00 & 0.00 & 0.00 & 0.00 & 0.00 & 2.00 & 2.00 & 2.00 & 2.00 & 0.00 & 5.98 & 1.36 & 0.00 & 5.68 \\ 
n=2000 & 2.00 & 0.00 & 0.00 & 0.00 & 0.00 & 0.00 & 0.06 & 0.12 & 0.10 & 0.00 & 0.00 & 0.00 & 0.02 & 0.00 & 0.00 & 2.00 & 2.00 & 2.00 & 2.00 & 0.00 & 5.98 & 1.68 & 0.00 & 5.72 \\ 
n=5000 & 2.00 & 0.00 & 0.00 & 0.00 & 0.00 & 0.00 & 0.00 & 0.20 & 0.20 & 0.00 & 0.00 & 0.00 & 0.00 & 0.00 & 0.00 & 2.00 & 2.00 & 2.00 & 2.00 & 0.00 & 5.98 & 1.60 & 0.00 & 5.76 \\ 
n=10000 & 2.00 & 0.00 & 0.00 & 0.00 & 0.00 & 0.00 & 0.02 & 0.26 & 0.22 & 0.00 & 0.00 & 0.00 & 0.00 & 0.00 & 0.00 & 2.00 & 2.00 & 2.00 & 2.00 & 0.00 & 5.98 & 1.58 & 0.00 & 5.88 \\ 
  \midrule
 & \multicolumn{24}{c}{K=5} \\ \midrule
Sample size & BHI & CHI & DI & GDI11 & GDI12 & GDI13 & GDI21 & GDI22 & GDI23 & GDI31 & GDI32 & GDI33 & GDI41 & GDI42 & GDI43 & GDI51 & GDI52 & GDI53 & KWI & PBMI & PBI & RLI & SI & TWBI \\  \midrule
n=300 & 3.00 & 0.00 & 2.22 & 2.22 & 3.00 & 2.82 & 1.50 & 2.98 & 2.62 & 1.26 & 3.00 & 2.76 & 1.56 & 3.00 & 2.94 & 2.98 & 2.98 & 2.96 & 2.98 & 1.06 & 5.00 & 3.00 & 3.00 & 4.06 \\ 
n=500 & 3.00 & 0.00 & 2.82 & 2.82 & 3.00 & 3.00 & 2.28 & 3.00 & 2.76 & 2.34 & 3.00 & 2.82 & 2.58 & 3.00 & 3.00 & 3.02 & 3.00 & 3.00 & 3.00 & 1.20 & 5.00 & 3.00 & 3.00 & 4.72 \\ 
n=1000 & 3.00 & 0.00 & 2.94 & 2.94 & 3.00 & 3.00 & 2.52 & 3.00 & 2.46 & 2.88 & 3.00 & 3.00 & 2.88 & 3.00 & 3.00 & 3.04 & 3.00 & 3.00 & 2.98 & 1.64 & 5.00 & 3.00 & 3.00 & 4.70 \\ 
n=2000 & 3.00 & 0.00 & 3.00 & 3.00 & 3.00 & 3.00 & 2.52 & 3.00 & 2.52 & 2.94 & 3.00 & 3.00 & 2.94 & 3.00 & 3.00 & 3.00 & 3.00 & 3.00 & 3.00 & 1.76 & 5.00 & 3.00 & 3.00 & 4.56 \\ 
n=5000 & 3.00 & 0.00 & 3.00 & 3.00 & 3.00 & 3.00 & 2.64 & 3.00 & 2.46 & 3.00 & 3.00 & 3.00 & 3.00 & 3.00 & 3.00 & 3.00 & 3.00 & 3.00 & 3.00 & 1.96 & 5.00 & 3.00 & 3.00 & 4.72 \\ 
n=10000 & 3.00 & 0.00 & 3.00 & 3.00 & 3.00 & 3.00 & 2.88 & 3.00 & 2.04 & 3.00 & 3.00 & 3.00 & 3.00 & 3.00 & 3.00 & 3.00 & 3.00 & 3.00 & 3.00 & 1.92 & 5.00 & 3.00 & 3.00 & 4.70 \\ 
  \midrule
 & \multicolumn{24}{c}{K=6} \\ \midrule
Sample size & BHI & CHI & DI & GDI11 & GDI12 & GDI13 & GDI21 & GDI22 & GDI23 & GDI31 & GDI32 & GDI33 & GDI41 & GDI42 & GDI43 & GDI51 & GDI52 & GDI53 & KWI & PBMI & PBI & RLI & SI & TWBI \\  \midrule
n=300 & 4.00 & 0.00 & 0.04 & 0.04 & 0.16 & 0.16 & 2.80 & 4.00 & 4.00 & 0.00 & 2.24 & 0.56 & 0.00 & 2.08 & 0.56 & 4.00 & 3.96 & 4.00 & 4.00 & 0.00 & 4.00 & 4.00 & 1.52 & 3.68 \\ 
n=500 & 4.00 & 0.00 & 0.00 & 0.00 & 0.00 & 0.00 & 2.72 & 4.00 & 4.00 & 0.00 & 2.24 & 0.32 & 0.00 & 1.92 & 0.24 & 4.00 & 3.98 & 4.00 & 4.00 & 0.00 & 4.00 & 4.00 & 1.92 & 3.66 \\ 
n=1000 & 4.00 & 0.00 & 0.02 & 0.02 & 0.24 & 0.24 & 3.44 & 4.00 & 4.00 & 0.00 & 2.00 & 0.40 & 0.00 & 1.84 & 0.40 & 4.00 & 3.96 & 4.00 & 4.00 & 0.00 & 4.00 & 4.00 & 1.84 & 3.66 \\ 
n=2000 & 4.00 & 0.00 & 0.02 & 0.02 & 0.26 & 0.08 & 3.60 & 4.00 & 4.00 & 0.00 & 1.60 & 0.00 & 0.00 & 1.44 & 0.00 & 4.00 & 4.00 & 4.00 & 4.00 & 0.00 & 4.00 & 4.00 & 1.12 & 3.70 \\ 
n=5000 & 4.00 & 0.00 & 0.08 & 0.08 & 0.32 & 0.24 & 4.00 & 4.00 & 4.00 & 0.00 & 1.68 & 0.00 & 0.00 & 1.28 & 0.00 & 4.00 & 4.00 & 4.00 & 4.00 & 0.00 & 4.00 & 4.00 & 1.36 & 3.80 \\ 
n=10000 & 4.00 & 0.00 & 0.00 & 0.00 & 0.40 & 0.32 & 4.00 & 4.00 & 4.00 & 0.00 & 1.20 & 0.00 & 0.00 & 1.44 & 0.00 & 4.00 & 4.00 & 4.00 & 4.00 & 0.00 & 4.00 & 4.00 & 0.88 & 3.80 \\ 
\bottomrule
\end{tabular}
\end{small}
\end{sidewaystable}

Table \ref{gmmp} contains the results of the $\alpha$--GPCM for all dimensionalities. 

\begin{table}
\centering
\caption{Results of the $\alpha$--GPCM. Absolute distance between the estimated and the true number of clusters.}
\label{gmmp}
\begin{tabular}{l|rrrr|rrrr|rrrr}
\toprule
& \multicolumn{4}{c}{p=3} & \multicolumn{4}{c}{p=5} & \multicolumn{4}{c}{p=10}\\
\midrule
Sample size & K=3 & K=4 & K=5 & K=6 & K=3 & K=4 & K=5 & K=6 & K=3 & K=4 & K=5 & K=6 \\  \midrule
n=300   & 0.17 & 0.32 & 0.57 & 0.77 & 0.05 & 0.34 & 0.37 & 0.21 & 0.00 & 0.02 & 0.70 & 0.12 \\ 
n=500   & 0.00 & 0.17 & 0.53 & 1.16 & 0.02 & 0.22 & 0.36 & 0.47 & 0.00 & 0.00 & 0.82 & 0.58 \\ 
n=1000  & 1.14 & 0.57 & 0.81 & 0.99 & 0.78 & 0.27 & 0.51 & 0.49 & 0.00 & 0.10 & 0.36 & 0.60 \\ 
n=2000  & 0.04 & 0.59 & 1.38 & 1.54 & 0.15 & 0.55 & 0.86 & 0.29 & 0.40 & 0.34 & 1.06 & 0.34 \\ 
n=5000  & 1.45 & 0.77 & 2.37 & 1.79 & 1.15 & 0.60 & 0.98 & 0.55 & 0.40 & 1.60 & 1.36 & 0.14 \\ 
n=10000 & 2.12 & 1.96 & 2.81 & 2.40 & 4.06 & 1.65 & 3.18 & 1.13 & 2.78 & 3.66 & 1.92 & 0.14 \\ \midrule
\bottomrule
\end{tabular}
\end{table}

Table \ref{time} presents the ratio of of the computational cost between the $\alpha$--GPCM and the $\alpha$--$K$--means when $p=3$\footnote{The computational cost ratios are similarwhen $p=5$ and $p=10$ and hence not presented.}. Regardless of the number of components of the compositional data, $\alpha$--$K$--means is always faster than $\alpha$--GPCM, with a range of 40 up to 600 times faster. 

\begin{table}[ht]
\centering
\caption{Computational cost ratio of $\alpha$--GPCM relative to $\alpha$--$K$--means when $p=3$.}
\label{time}
\begin{tabular}{l|rrrr}
\toprule
Sample size & K=3 & K=4 & K=5 & K=6 \\ \midrule
n=300    & 495.29 & 353.87 & 399.67 & 247.83 \\ 
n=500    & 573.83 & 534.27 & 304.42 & 280.91 \\ 
n=1,000  & 605.56 & 512.26 & 376.35 & 319.88 \\ 
n=2,000  & 438.52 & 373.94 & 320.63 & 253.79 \\ 
n=5,000  & 169.92 & 152.12 & 121.16 & 102.77 \\ 
n=10,000 & 79.56 & 67.64 & 60.50 & 47.29 \\ \bottomrule 
\end{tabular}
\end{table}

\section{Real data analysis}  \label{sec:real}
A real dataset will be used to illustrate the performance of the $\alpha$--$K$--means and of the $\alpha$--GPCM. The Honey dataset consists of 429 measurements on twelve mineral elements. There are 201 observations with pure honey, 183 observations were coming from adulterated honey, that is pure honey with 5\% high fructose starch syrup added, and 45 observations coming from syrup. The peculiarity of this dataset is that it contained any zero observations, hence only strictly positive values of $\alpha$ were feasible. 

The minimum valued and maximum valued CVIs computed for the $\alpha$--$K$--means are presented in Table \ref{real:cvi}, respectively. The DBI, RTI and XBI agree that 2 clusters should be selected, with $\alpha=0.1$. With regards to the maximum valued CVIs, most of them agree with the choice of 2 clusters, but at various values of $\alpha$. The optimal model for the $\alpha$--GPCM was the VVE, with $\alpha=1$, which chose 8 clusters, thus performing very poorly. A closer examination of the results of the $\alpha$--$K$--means with $\alpha=0.1$ revealed the two clusters form a reasonable solution. The first group identified contains the pure and adulterated honey observations and the second group contains the syrup observations. Regarding the computational cost, the $\alpha$--$K$--means required 0.27 seconds, while the $\alpha$--GPCM required 115.51 seconds, indicating that the latter approach is more than 400 times slower. 

\begin{figure}[!ht]
\centering
\begin{tabular}{c}
\includegraphics[scale = 0.6]{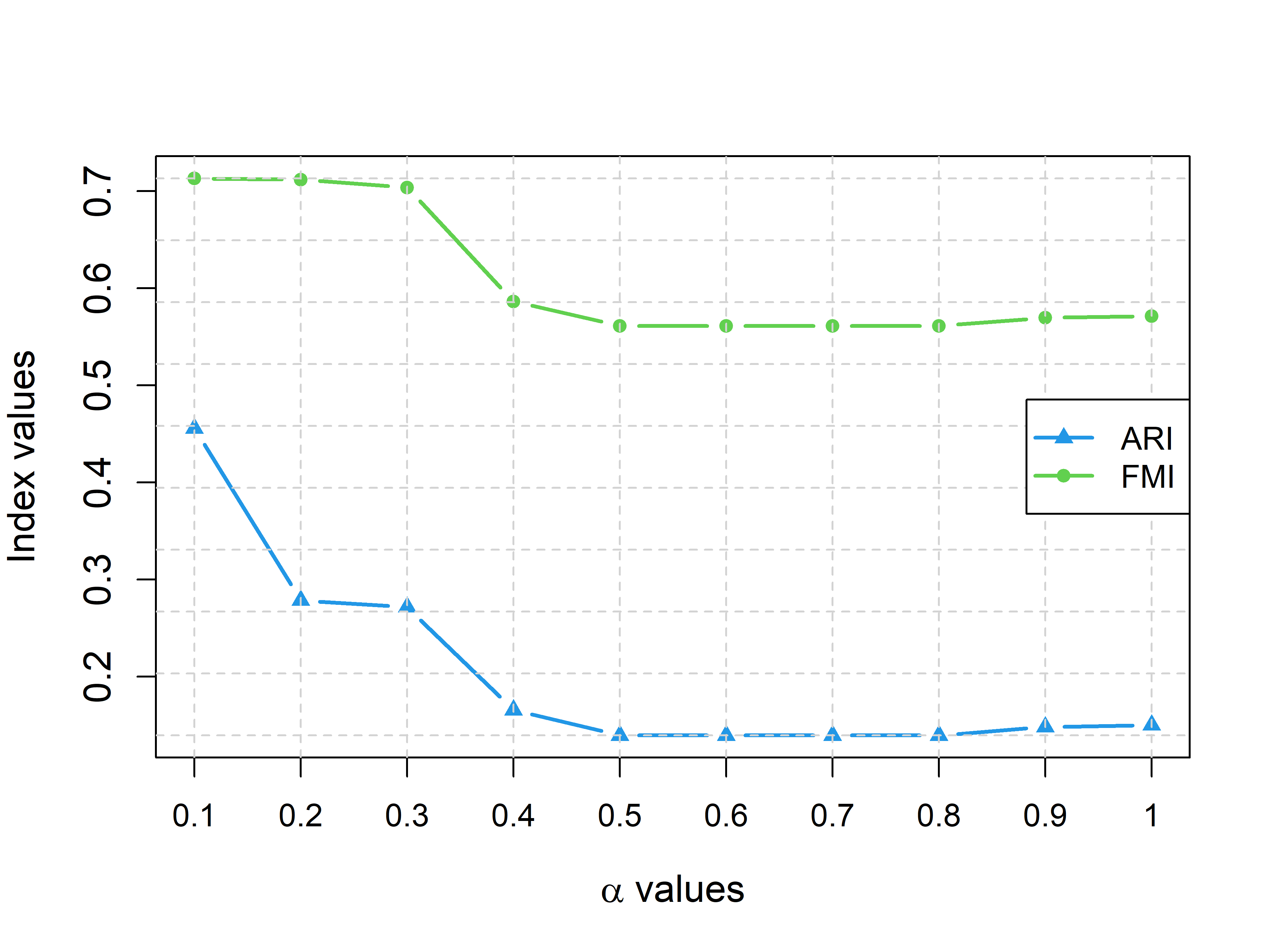} \\
(a) ARI and Folkes-Mallows indices for the $\alpha$--$K$--means. \\
\includegraphics[scale = 0.6]{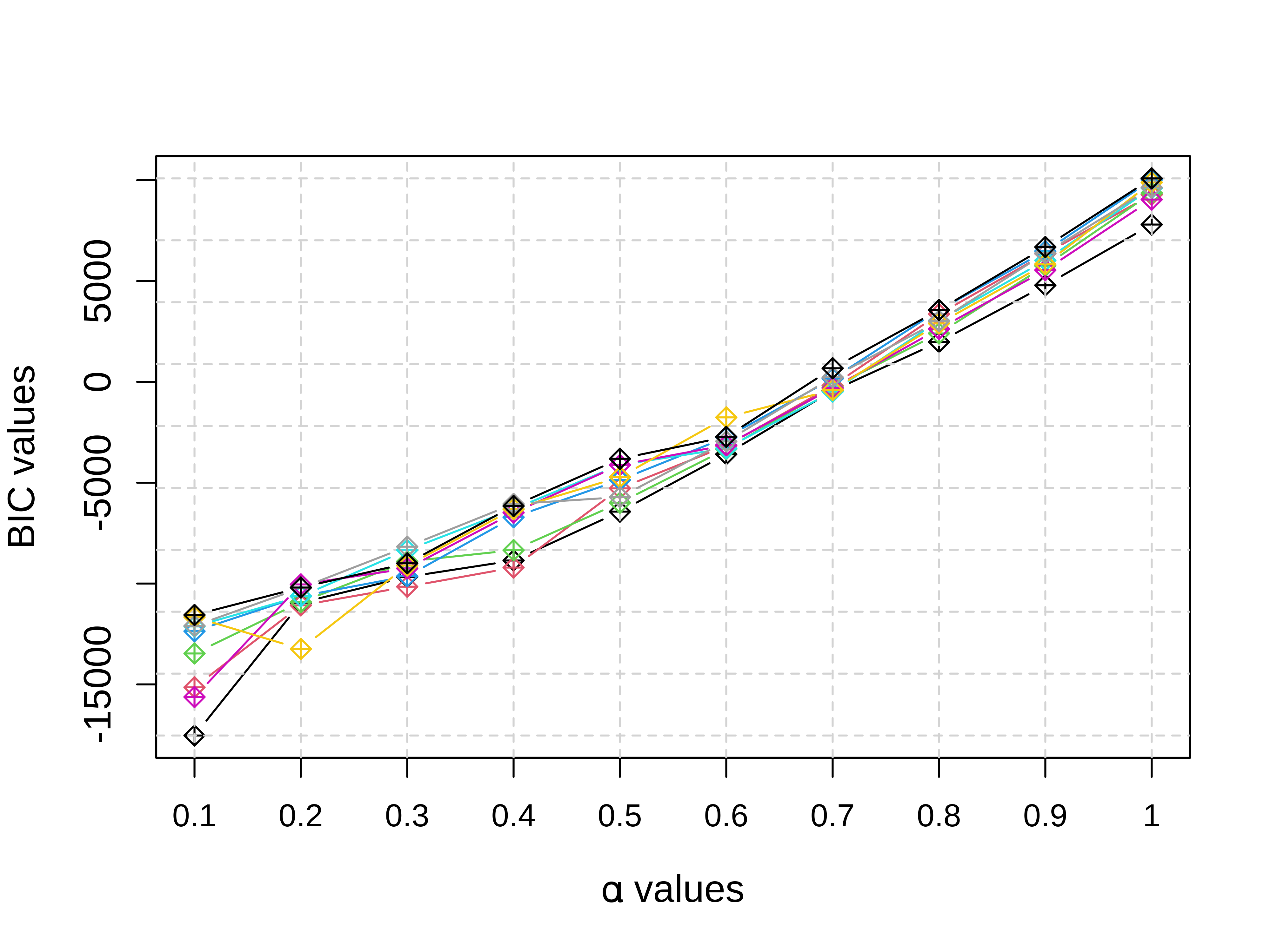} \\
(b) BIC of the $\alpha$--GMM. 
\end{tabular}
\caption{Sensitivity of the results with regards to the $\alpha$-values for the Honey dataset.}
\label{arifm1}
\end{figure}

The choice of the value of $\alpha$ seems to have an impact on the number of clusters. To illustrate this effect we computed the adjusted Rand index (ARI) \citep{hubert1985} and the Fowlkes-Mallows index (FMI) \citep{fowlkes1983} when we perform the $\alpha$--$K$--means and the $\alpha$--GPCM using two clusters for a range of values of $\alpha$. Figure \ref{arifm1}1(a) shows that the highest values of ARI and FMI are observed when $\alpha=0.1$, and the same conclusion is drawn when three clusters were used. Figure \ref{arifm1}(b) shows the (negated) BIC of the $\alpha$--GPCM, where we can see that the optimal value of $\alpha$ is 1. 

\begin{table}[ht]
\centering
\caption{CVIs for the Honey dataset.}
\label{real:cvi}
\begin{tabular}{lrrr|lrrr}
\toprule
\multicolumn{4}{c}{Maximum valued CVIs} & \multicolumn{4}{c}{Minimum valued CVIs} \\
\midrule
Index & $\alpha$ & K & Value & Index & $\alpha$ & K & Value \\
\midrule
  BHI & 1.00 & 2 & 10.88  &  BRI & 0.90 & 5 & 322.96 \\ 
  CHI & 0.90 & 3 & 285.11 &  DBI & 0.10 & 2 & 0.80 \\  
  Dunn & 0.10 & 2 & 0.58  &  DRI & 1.00 & 2 & 5.54 \\
  GDI11 & 0.10 & 2 & 0.58 &  LDRI & 1.00 & 2 & 734.30 \\  
  GDI12 & 0.10 & 2 & 2.71 &  LSSI & 1.00 & 2 & -0.57 \\ 
  GDI13 & 0.10 & 2 & 0.91 &  MRI & 1.00 & 4 & 0.32 \\ 
  GDI21 & 0.90 & 2 & 1.30 &  RTI & 0.10 & 2 & 0.15 \\ 
  GDI22 & 0.90 & 2 & 6.53 &  SSI & 0.10 & 5 & -20464.70 \\  
  GDI23 & 0.90 & 2 & 2.34 &  XBI & 0.10 & 2 & 0.20 \\  
  GDI31 & 0.50 & 2 & 0.83 & \\ 
  GDI32 & 0.10 & 2 & 3.69 & \\ 
  GDI33 & 0.40 & 2 & 1.29 & \\ 
  GDI41 & 0.50 & 2 & 0.71 & \\ 
  GDI42 & 0.10 & 2 & 3.15 & \\ 
  GDI43 & 0.40 & 2 & 1.10 & \\ 
  GDI51 & 0.50 & 2 & 0.26 & \\ 
  GDI52 & 0.40 & 2 & 1.16 & \\ 
  GDI53 & 0.40 & 2 & 0.42 & \\ 
  KWI & 0.30 & 2 & 2.108e+25 & \\ 
  PBMI & 0.10 & 2 & 16.99 & \\ 
  PBI & 0.30 & 6 & -1.07 & \\ 
  RLI & 0.10 & 2 & 0.44 & \\ 
  SI & 0.10 & 2 & 0.54 & \\ 
  TBI & 0.10 & 5 & 125.06 & \\ 
\bottomrule
\end{tabular}
\end{table}

\clearpage
\section{Conclusions}  \label{Sec:conc}
We introduced two novel simplicial clustering methodologies that incorporate flexibility through the application of the non-parametric $\alpha$--$K$--means algorithm and the parametric $\alpha$--GPCM model. This flexibility arises from the $\alpha$--transformation, which is applied prior to clustering. In contrast to log-ratio based approaches, which are invalid in the presence of zero values, the $\alpha$--transformation naturally accommodates such cases. The choice of the tuning parameter $\alpha$ was addressed using cluster validity indices (CVIs), commonly employed for determining the number of clusters in $K$--means, and the BIC, typically adopted for GPCMs.

The simulation experiments highlighted a limitation of the CVIs, namely their frequent inability to correctly identify the true number of clusters, even under scenarios where the clusters were well separated. Moreover, while $\alpha$--GPCMs demonstrated reasonable performance when applied to compositional data, the $\alpha$--$K$--means based approach generally outperformed them, particularly with respect to correctly recovering the number of clusters and, more importantly, computational efficiency. Analysis of a real-world dataset further underscored the shortcomings of the $\alpha$--GPCM model.

We have created an \textit{R} package called \textsf{CompositionalClust} \citep{compositionalclust2025} to apply all the aforementioned algorithms, that is available to download from \href{https://cran.r-project.org/web/packages/CompositionalClust/index.html}{CRAN}.

Future work should be done on the directions of alternative transformations and or mixture models. Exploration of non-linear clustering algorithms, such as the Density-Based Spatial Clustering of Applications with Noise \citep{ester1996} is another option. 

\clearpage
\section*{Appendix}
\renewcommand{\theequation}{A.\arabic{equation}}
\renewcommand\thefigure{A.\arabic{figure}}    
\renewcommand\thetable{A.\arabic{table}}    
\setcounter{equation}{0}  
\setcounter{figure}{0}  
\setcounter{table}{0}  

\begin{figure}[!ht]
\centering
\begin{tabular}{cc}
\includegraphics[scale = 0.55, trim = 50 70 0 0]{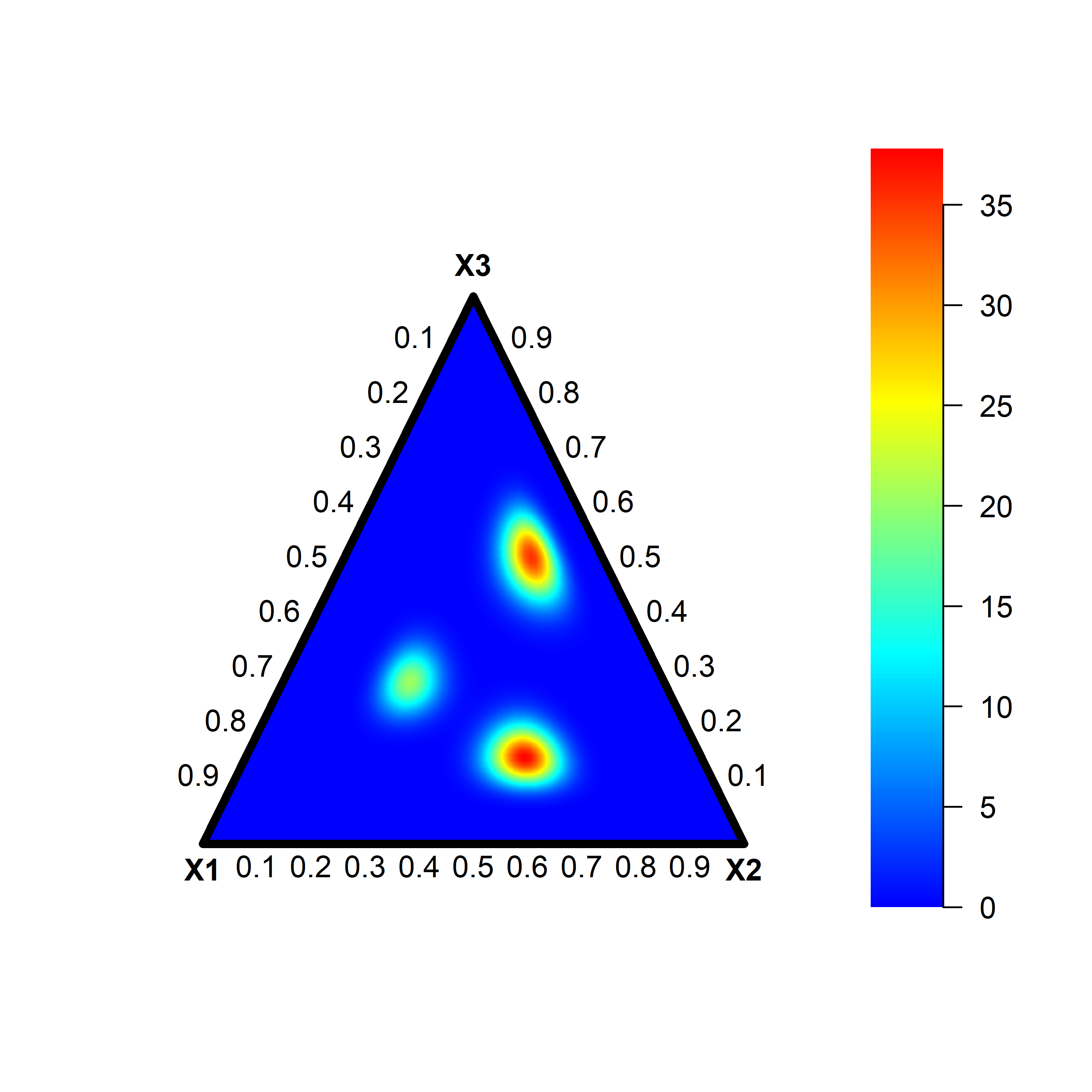} & 
\includegraphics[scale = 0.55, trim = 60 70 0 0]{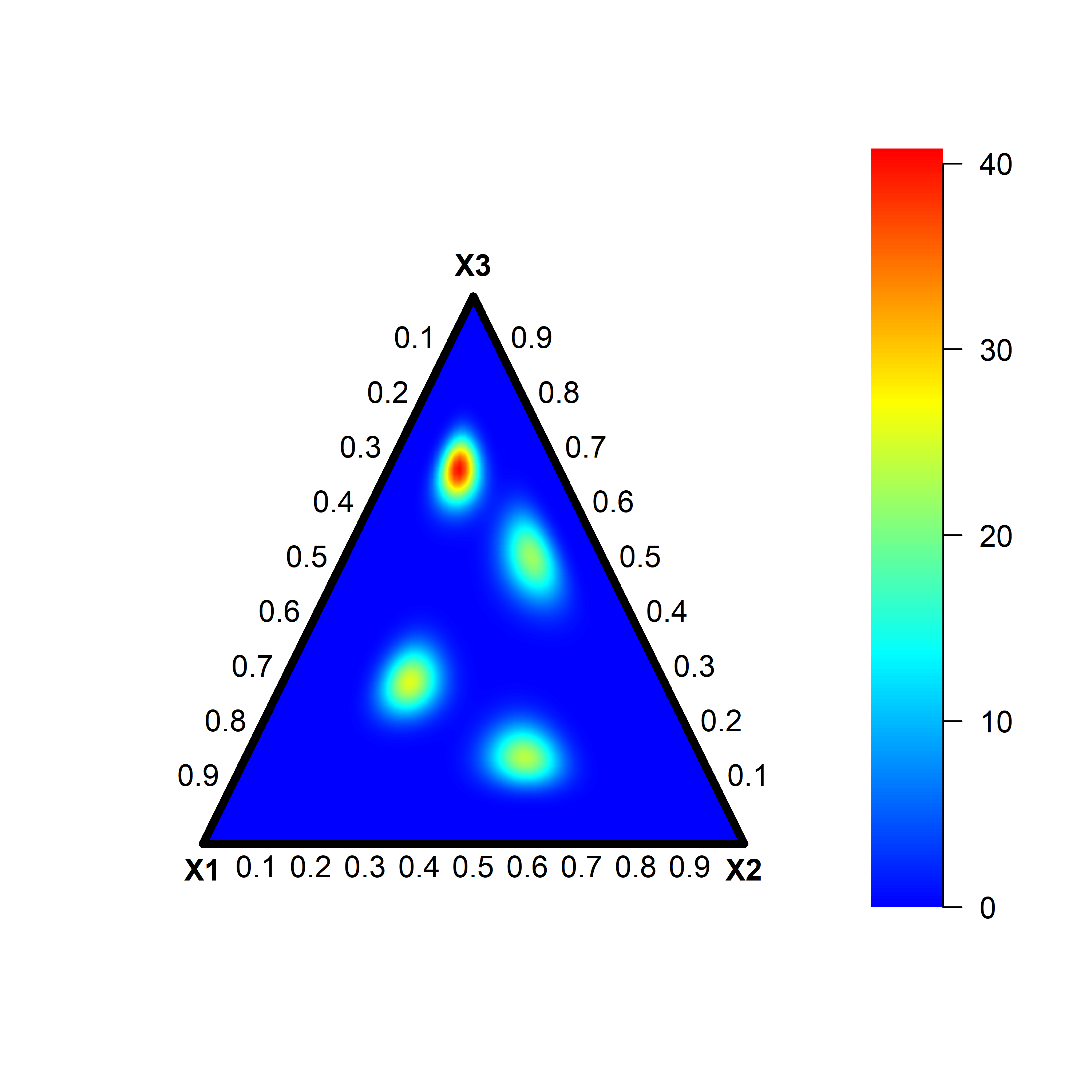} \\
(a) & (b) \\
\includegraphics[scale = 0.55, trim = 50 70 0 0]{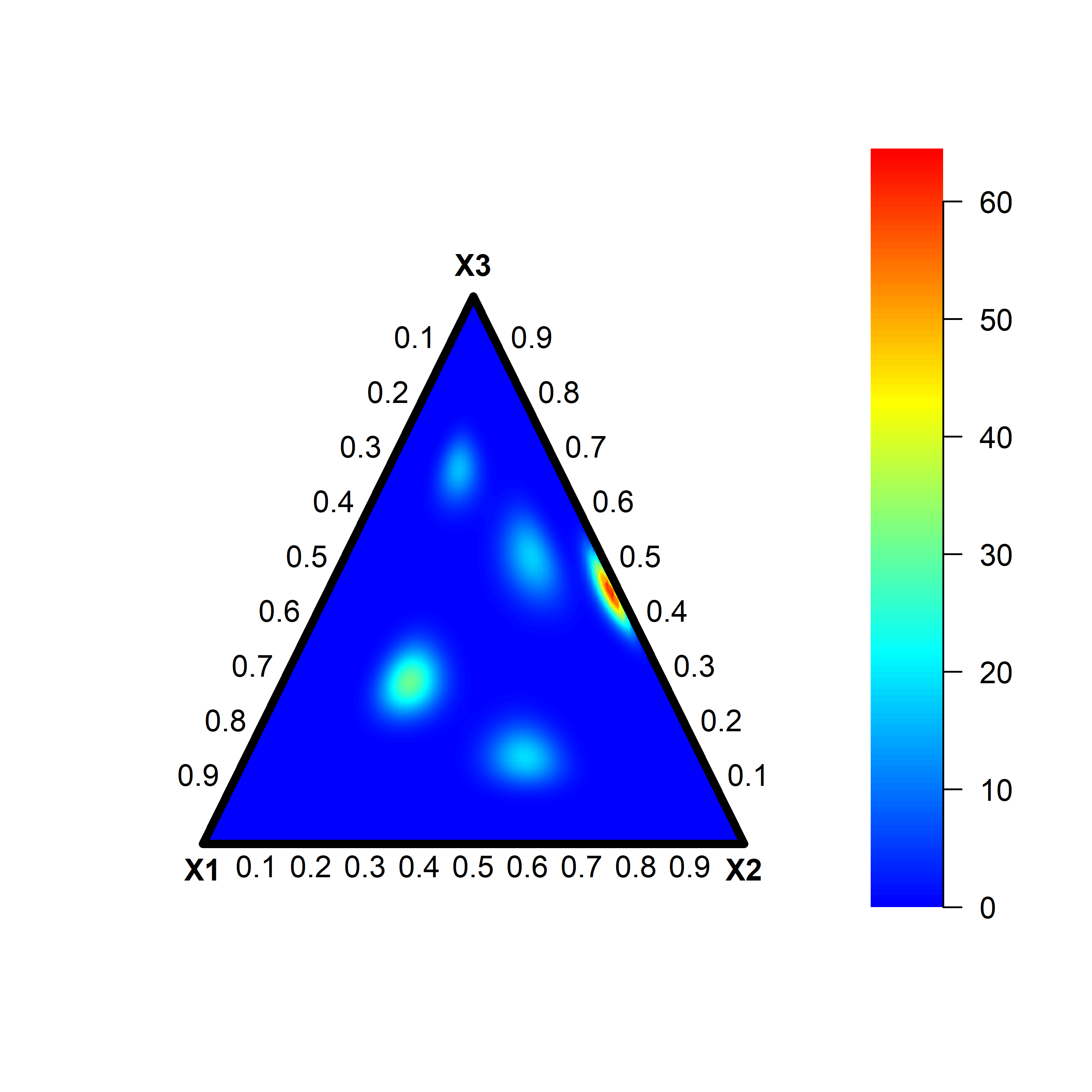} & 
\includegraphics[scale = 0.55, trim = 60 70 0 0]{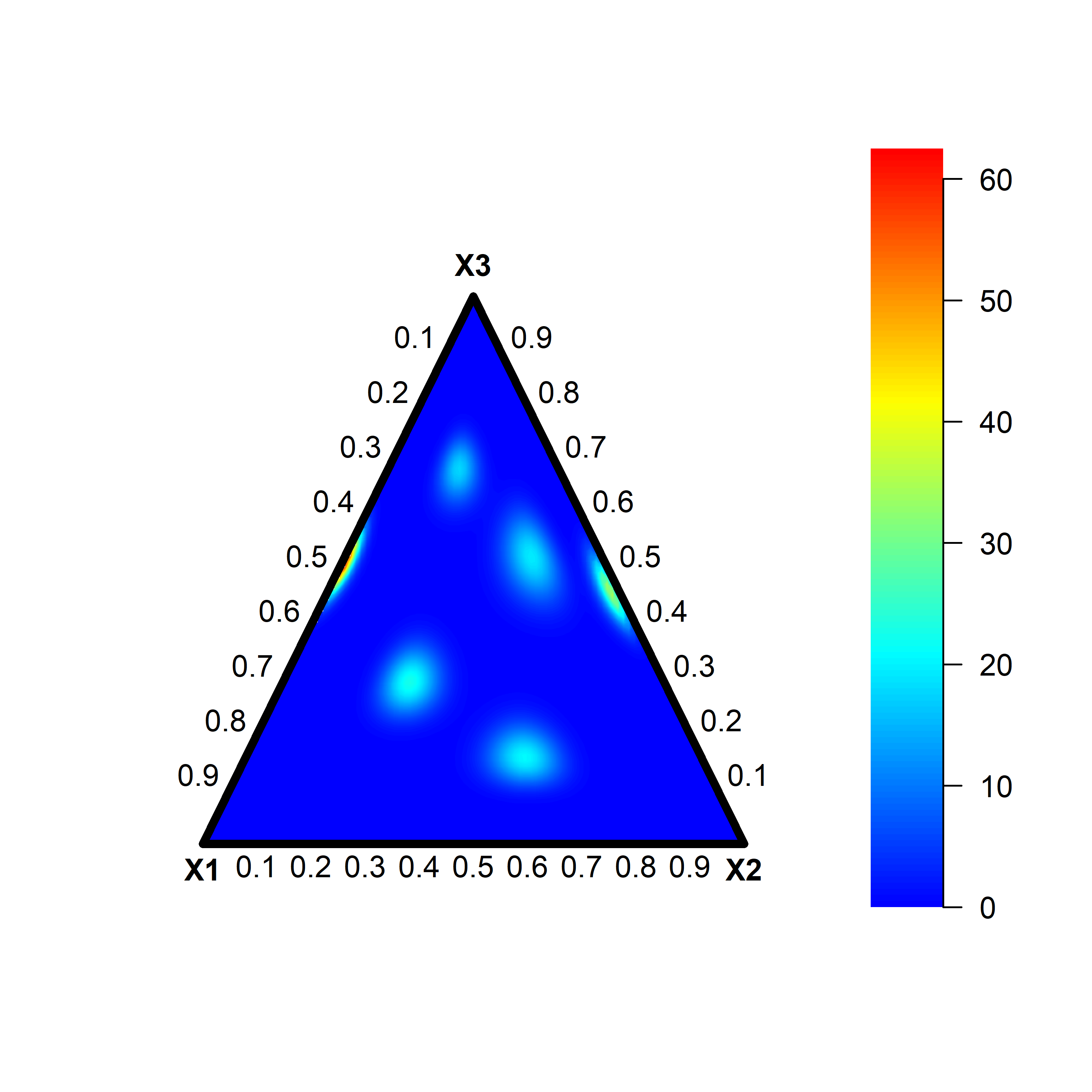} \\
(c) & (d) \\
\end{tabular}
\caption{Contour plots of the 4 DMMS employed in the simulation studies when $p=3$.}
\label{cont.dmm}
\end{figure}

The density function of the Dirichlet function is $f\left(\bm{y}; \bm{a} \right)=\frac{\Gamma\left(\sum_{i=1}^p\phi a_i\right)}{\prod_{i=1}^p\Gamma\left(\phi a_i\right)}\prod_{i=1}^py_i^{\phi a_i-1}$. In our simulation studies we generated data using a DMM model whose density is given by $f\left(\bm{y};\bm{a},\bm{\pi}\right)=\sum_{j=1}^K\pi_jf\left(\bm{y};\bm{a}_j\right)$. Tables \ref{dmmp3}-\ref{dmmp5} contain the specifications of the DMMs used in the simulation studies, while Figure \ref{cont.dmm} visualizes the DMM when $p=3$.

\begin{table}[ht]
\centering
\caption{The parameters of the 4 DMMs used in the simulation studies when $p=3$.}
\label{dmmp3}
\begin{tabular}{l|rrr|c}
\toprule
& \multicolumn{3}{c}{Parameters} & Cluster probabilities \\  \midrule
DMM 1 & 12.00 & 30.00 & 45.00 & 0.40 \\ 
      & 32.00 & 50.00 & 16.00 & 0.40 \\ 
      & 55.00 & 28.00 & 35.00 & 0.20 \\  \midrule
DMM 2 & 12.00 & 30.00 & 45.00 & 0.30 \\ 
      & 25.00 & 18.00 & 90.00 & 0.30 \\ 
      & 55.00 & 28.00 & 35.00 & 0.20 \\ 
      & 32.00 & 50.00 & 16.00 & 0.20 \\  \midrule
DMM 3 & 12.00 & 30.00 & 45.00 & 0.20 \\ 
      & 25.00 & 18.00 & 90.00 & 0.10 \\ 
      & 55.00 & 28.00 & 35.00 & 0.30 \\ 
      & 32.00 & 50.00 & 16.00 & 0.20 \\ 
      & 3.00 & 68.00 & 60.00 & 0.20 \\  \midrule
DMM 4 & 12.00 & 30.00 & 45.00 & 0.20 \\ 
      & 32.00 & 50.00 & 16.00 & 0.24 \\ 
      & 55.00 & 28.00 & 35.00 & 0.21 \\ 
      & 3.00 & 68.00 & 60.00 & 0.11 \\ 
      & 25.00 & 18.00 & 90.00 & 0.13 \\ 
      & 75.00 & 2.00 & 80.00 & 0.11 \\  \bottomrule 
\end{tabular}
\end{table}

\begin{table}[ht]
\centering
\caption{The parameters of the 4 DMMs used in the simulation studies when $p=5$.}
\label{dmmp5}
\begin{tabular}{l|rrrrr|c}
\toprule
& \multicolumn{5}{c}{Parameters} & Cluster probabilities \\  \midrule
DMM 1 & 12.00 & 30.00 & 45.00 & 30.00 & 12.00 & 0.30 \\ 
      & 32.00 & 50.00 & 16.00 & 50.00 & 32.00 & 0.30 \\ 
      & 55.00 & 28.00 & 35.00 & 28.00 & 55.00 & 0.40 \\  \midrule
DMM 2 & 12.00 & 30.00 & 45.00 & 30.00 & 12.00 & 0.25 \\ 
      & 25.00 & 18.00 & 90.00 & 18.00 & 25.00 & 0.25 \\ 
      & 55.00 & 28.00 & 35.00 & 28.00 & 55.00 & 0.25 \\ 
      & 32.00 & 50.00 & 16.00 & 50.00 & 32.00 & 0.25 \\  \midrule      
DMM 3 & 12.00 & 30.00 & 45.00 & 30.00 & 12.00 & 0.35 \\ 
      & 25.00 & 18.00 & 90.00 & 18.00 & 25.00 & 0.15 \\ 
      & 55.00 & 28.00 & 35.00 & 28.00 & 55.00 & 0.25 \\ 
      & 32.00 & 50.00 & 16.00 & 50.00 & 32.00 & 0.10 \\ 
      & 3.00  & 68.00 & 60.00 & 68.00 & 3.00  & 0.15 \\  \midrule
DMM 4 & 12.00 & 30.00 & 45.00 & 30.00 & 12.00 & 0.22 \\ 
      & 32.00 & 50.00 & 16.00 & 50.00 & 32.00 & 0.22 \\ 
      & 55.00 & 28.00 & 35.00 & 28.00 & 55.00 & 0.19 \\ 
      & 3.00  & 68.00 & 60.00 & 68.00 & 3.00  & 0.11 \\ 
      & 25.00 & 18.00 & 90.00 & 18.00 & 25.00 & 0.13 \\ 
      & 75.00 & 2.00  & 80.00 & 2.00  & 75.00 & 0.13 \\  \bottomrule
\end{tabular}
\end{table}

\begin{table}[ht]
\centering
\caption{The parameters of the 4 DMMs used in the simulation studies when $p=10$.}
\label{dmmp10}
\begin{tabular}{l|rrrrrrrrrr|c}
\toprule
& \multicolumn{10}{c}{Parameters} & Cluster probabilities \\  \midrule
DMM 1 & 12.00 & 30.00 & 45.00 & 30.00 & 12.00 & 17.00 & 31.00 & 42.00 & 30.00 & 8.00 & 0.30 \\ 
      & 32.00 & 50.00 & 16.00 & 50.00 & 32.00 & 34.00 & 53.00 & 15.00 & 46.00 & 32.00 & 0.30 \\ 
      & 55.00 & 28.00 & 35.00 & 28.00 & 55.00 & 55.00 & 32.00 & 39.00 & 25.00 & 57.00 & 0.40 \\  \midrule 
DMM 2 & 12.00 & 30.00 & 45.00 & 30.00 & 12.00 & 13.00 & 25.00 & 44.00 & 34.00 & 8.00 & 0.25 \\ 
      & 25.00 & 18.00 & 90.00 & 18.00 & 25.00 & 23.00 & 17.00 & 93.00 & 14.00 & 25.00 & 0.25 \\ 
      & 55.00 & 28.00 & 35.00 & 28.00 & 55.00 & 59.00 & 28.00 & 39.00 & 27.00 & 53.00 & 0.25 \\ 
      & 32.00 & 50.00 & 16.00 & 50.00 & 32.00 & 28.00 & 53.00 & 21.00 & 46.00 & 31.00 & 0.25 \\  \midrule
DMM 3 & 12.00 & 30.00 & 45.00 & 30.00 & 12.00 & 11.00 & 31.00 & 47.00 & 31.00 & 16.00 & 0.35 \\ 
      & 25.00 & 18.00 & 90.00 & 18.00 & 25.00 & 23.00 & 16.00 & 86.00 & 23.00 & 29.00 & 0.15 \\ 
      & 55.00 & 28.00 & 35.00 & 28.00 & 55.00 & 52.00 & 31.00 & 31.00 & 24.00 & 59.00 & 0.25 \\ 
      & 32.00 & 50.00 & 16.00 & 50.00 & 32.00 & 36.00 & 47.00 & 17.00 & 54.00 & 35.00 & 0.10 \\ 
      & 3.00 & 68.00 & 60.00 & 68.00 & 3.00 & 1.00 & 69.00 & 62.00 & 65.00 & -1.00 & 0.15 \\  \midrule
DMM 4 & 12.00 & 30.00 & 45.00 & 30.00 & 12.00 & 9.00 & 31.00 & 40.00 & 34.00 & 15.00 & 0.22 \\ 
      & 32.00 & 50.00 & 16.00 & 50.00 & 32.00 & 37.00 & 45.00 & 11.00 & 50.00 & 32.00 & 0.22 \\ 
      & 55.00 & 28.00 & 35.00 & 28.00 & 55.00 & 54.00 & 26.00 & 40.00 & 33.00 & 60.00 & 0.19 \\ 
      & 3.00 & 68.00 & 60.00 & 68.00 & 3.00 & 1.00 & 71.00 & 61.00 & 65.00 & 6.00 & 0.11 \\ 
      & 25.00 & 18.00 & 90.00 & 18.00 & 25.00 & 28.00 & 18.00 & 90.00 & 22.00 & 23.00 & 0.13 \\ 
      & 75.00 & 2.00 & 80.00 & 2.00 & 75.00 & 72.00 & -1.00 & 76.00 & 3.00 & 77.00 & 0.13 \\ \bottomrule
\end{tabular}
\end{table}

\clearpage
\bibliographystyle{chicago}
\bibliography{ref}

\end{document}